\documentclass[]{article}

\usepackage{graphicx}

\usepackage{tikz}
\usepackage{caption}
\usepackage{epic,eepic}

\usepackage{rotating}
\usepackage{booktabs}
\usepackage{multirow}

\usepackage{amssymb}

\newtheorem{rem}{Remark}[section]
\newcommand{\br}{\begin{rem}}
\newcommand{\er}{\end{rem}}
\newtheorem{ex}{Example}[section]
\newcommand{\bex}{\begin{ex}}
\newcommand{\eex}{\end{ex}}
\newtheorem{Def}{Definition}[section]
\newcommand{\bd}{\begin{Def}}
\newcommand{\ed}{\end{Def}}
\newtheorem{theorem}{Theorem}[section]
\newcommand{\bt}{\begin{theorem}}
\newcommand{\et}{\end{theorem}}
\newtheorem{corollary}{Corollary}[section]
\newcommand{\bc}{\begin{corollary}}
\newcommand{\ec}{\end{corollary}}
\newtheorem{lemma}{Lemma}[section]
\newcommand{\bl}{\begin{lemma}}
\newcommand{\el}{\end{lemma}}
\newcommand{\be}{\begin{equation}}
\newcommand{\ee}{\end{equation}}
\newcommand{\bea}{\begin{eqnarray}}
\newcommand{\eea}{\end{eqnarray}}

\newtheorem{prop}{Proposition}[section]
\newcommand{\bpr}{\begin{prop}}
\newcommand{\epr}{\end{prop}}
\newcommand{\dis}{\displaystyle}
\newtheorem{proof}{Proof}[section]
\newcommand{\bpf}{\begin{proof}}
\newcommand{\epf}{\end{proof}}

\usepackage{tikz-3dplot}
\usepackage{xifthen}
\usepackage{caption}

\tdplotsetmaincoords{60}{125}
\tdplotsetrotatedcoords{0}{20}{0} 

\usepackage[
labelfont=sf,
hypcap=false,
format=hang
]{caption}

\begin{document}

\title{Difference systems in bond and face variables and non-potential versions of discrete integrable systems}

\author{
Pavlos Kassotakis
\thanks{\emph{Present address:} Department of Mathematics, University of the Aegean, Karlovassi, 83200 Samos, Greece;
\newline \emph{e-mail:} {\tt  pavlos1978@gmail.com}}
\and
Maciej Nieszporski
\thanks{\emph{Present address:} Katedra Metod Matematycznych Fizyki, Wydzia\l{} Fizyki, Uniwersytet Warszawski,
ul. Pasteura 5, 02-093 Warszawa, Poland;
\emph{e-mail:} {\tt maciejun@fuw.edu.pl}}}

\maketitle

\begin{abstract}
\noindent Integrable discrete scalar equations defined on a~two or a three dimensional lattice can be rewritten as difference systems  in bond variables
or in face variables respectively. Both the difference systems in bond variables and the difference systems in face variables can be regarded as vector versions of the original equations. As a result, we link some of the discrete equations by difference substitutions and reveal the non-potential versions of  some consistent-around-the-cube  equations. We obtain higher-point configurations, including pairs of compatible six~points equations on the ${\mathbb Z}^2$ lattice together with associated seven points equations. Also we obtain a variety~of compatible ten point equations together with associated ten and twelve point equations on the ${\mathbb Z}^3$ lattice. Finally, we present integrable multiquadratic quad relations.

\smallskip
\noindent \textbf{Keywords:} Discrete integrable systems, difference substitutions, simplex equations
\end{abstract}

\section{Introduction}

 The aim of this  paper is a systematic investigation of links between discrete nonlinear integrable equations.
By a discrete nonlinear integrable equation we understand here such a recurrence that first, admits a B\"acklund transformation
that allows to build from a seed solution of the equation a family of solutions of the equation with at least one essential parameter,
second, possess Bianchi's type nonlinear superposition principle that allows us to superpose the solutions in a nonlinear way.
Discrete integrable systems (or their ultradiscrete versions) appear in various branches of mathematics and physics such as,
algebraic geometry \cite{Sakai}, cluster algebras \cite{Fomin1-2002,Fomin2-2002,Fordy-2014}, group theory
\cite{Kato2017}, difference geometry \cite{Sauer1970,BoSu-book,DOLIWA1997,KoSch}, statistical physics \cite{yang-1967,baxter-1970,Jimbo1990,drin,Kashaev1996,Sergeev-1998,Korepanov-1998,Bazhanov-2016}, just to name a few.


Integrable systems can appear in several disguises, this is one of the reasons why the link between integrable systems and the mentioned fields had been undiscovered for years. For instance,   {\it Menelaus of Alexandria theorem}, the {\it Desargues maps} and the  {\it star-triangle map } were recognized as essential in the theory of the integrable systems quite recently \cite{KoSch,Doliwa1177,Kashaev1996}.
A given integrable equation can be related to another by a point transformation, difference substitution or by a non-auto B\"acklund transformation \cite{KaNie1}.
We aim to reveal here some of such links. We focus here on difference substitutions. More precisely, we show that particular difference substitutions,
i.e. the ones that come from the mentioned systems in bond variables (or face variables), which we refer to as {\it bond systems} (or {\it face systems})   or the ones that come from the invariants of the introduced bond systems, see Sections \ref{secb} and \ref{sec4}, leads to difference systems which either are new and of the form that has not been studied systematically yet.


In particular we present the following forms of integrable equations
\begin{enumerate}
\item Pairs of compatible six point equations as non-potential versions of known integrable quad equations
\item Seven point equations associated with the pairs of the six point equations
\item A non-potential form of the Hirota-Miwa equation  that is referred to as the  gauge invariant form of the Hirota-Miwa equation \cite{Kuniba-1994}
\item  An asymmetric quad equation similar to the ones introduced in \cite{hyd-general,Adler_2011}
\item Multiquadratic quad relations
\item Pairs of compatible ten point equations
\item Ten point equations and twelve point equations associated with the pairs of the ten point equations
\end{enumerate}
Rewriting a given integrable equation defined on vertices of the ${\mathbb Z}^2$
(or ${\mathbb Z}^3$) lattice as a system of difference equations given on edges of the  corresponding ${\mathbb Z}^2$ lattice (or on faces  of the ${\mathbb Z}^3$  lattice)\cite{hi-via,Tasos}, opens a new perspective  on the theory of discrete integrable systems. As we shall see it serves as a unifying tool in the theory. In fact this type of systems appeared earlier in the integrable systems literature without referring them to as difference systems in bond or face variables. For instance the {\it first} and {\it second potentials} of  \cite{doliwa-santini} can be regarded as bond and face variables respectively.

To illustrate the idea (which we will refer here to as {\em vectorization} procedure) consider the lattice potential KdV equation \cite{NC,WaEs} (which also is referred to as $H1$)
\begin{equation}
\label{lpKdV}
\!\!\!\! (x_{m+1,n+1}-x_{m,n})(x_{m+1,n}-x_{m,n+1})=p_m-q_n,
\end{equation}
where $x$ is the dependent variable and $p_m$ and $q_n$ are given functions of the indicated (single) variable.
We can {\it vectorize} it by introducing the variables $(u,v)$
\[u_{m_1,n_1}=x_{m+1,n} x_{m,n}, \qquad v_{m_2,n_2}= x_{m,n+1} x_{m,n},\]
where we consider $u$ as a function  defined  on the set of horizontal edges
(i.e. pairs of vertices $\{(m,n),(m+1,n)\} =: (m_1,n_1) \in {\mathbb Z}^2 $)
and respectively $v$ as a function  defined  on set of vertical edges (i.e. pairs of vertices $\{(m,n),(m,n+1)\} =: (m_2,n_2) \in {\mathbb Z}^2$), see  Section~\ref{SETS} for  precise definitions.
Also  $p_m$ can be regarded as a function given on horizontal edges
and $q_n$ as a function given on vertical edges
i.e. we can write $p_{m_1}$ and $q_{n_2}$.
We get the following difference system in bond variables
\begin{equation}
\label{F4}
u_{m_1,n_1+1}=v_{m_2,n_2}\left(1+\frac{p_{m_1}-q_{n_2}}{u_{m_1,n_1}-v_{m_2,n_2}}\right) \quad v_{m_2+1,n_2}=u_{m_1,n_1}\left(1+\frac{p_{m_1}-q_{n_2}}{u_{m_1,n_1}-v_{m_2,n_2}}\right).
\end{equation}
We  thoroughly studied the path from lattice integrable systems to  systems in bond variables  and vice versa in  series of our papers \cite{KaNie,KaNie1,KaNie3}.

The goal of this article is to show that looking from the perspective of difference  systems in bond variables we can unify some  integrable equations and integrable relations (correspondences)
that so far were not connected.
We systematically study here two basic procedures. Namely,
\begin{itemize}
\item we rewrite the bond  systems   in terms of their  invariants with separated variables (see Section \ref{sec4})
e.g. in case of the system  (\ref{F4}), one of the underlying invariant is $\displaystyle H_{M,N}=\frac{u_{m_1,n_1}}{v_{m_2,n_2}}=\frac{x_{m+1,n}}{x_{m,n+1}},$ where the subscripts $M, N$ label anti-diagonals of the ${\mathbb Z}^2-$lattice as we describe it in Section \ref{SETS}.
The difference system (\ref{F4}) rewritten in terms of $H_{M,N},$  takes  form of a pair of compatible 6-point equations (\ref{QE-H}).
\item we rewrite the   bond systems in terms of the $u$ variable only by eliminating the variable $v.$ For example, in case of the system  (\ref{F4})
we get the following multiquadratic relation
\[\begin{array}{c}
{\displaystyle(u_{m_1+1,n_1} u_{m_1+1,n_1+1}-u_{m_1,n_1} u_{m_1,n_1+1})^2+ [p_{m_1+1}+u_{m_1+1,n_1}+u_{m_1+1,n_1+1}}-\\ [3mm]
 {\displaystyle (p_{m_1}+u_{m_1,n_1}+u_{m_1,n_1+1})]\cdot
[(p_{m_1+1}+u_{m_1+1,n_1}+u_{m_1+1,n_1+1}) u_{m_1,n_1} u_{m_1,n_1+1}} -\\ [3mm]
{\displaystyle (p_{m_1}+u_{m_1,n_1}+u_{m_1,n_1+1})u_{m_1+1,n_1} u_{m_1+1,n_1+1}
+ q_{n_2}(u_{m_1+1,n_1} u_{m_1+1,n_1+1}-}\\ [3mm]
{\displaystyle u_{m_1,n_1} u_{m_1,n_1+1})]=0.}
\end{array}\]
\end{itemize}
We also apply analogous procedures to integrable equations  on the ${\mathbb Z}^3$ lattice, revealing the structure of several ten and twelve point schemes. We are starting our paper with the introductory Sections \ref{SETS} and \ref{secb} and then we present our main results in Sections \ref{sec4} and \ref{sec5}. We are ending the article with some suggestions for further development in Section \ref{sec6}.

\section{The ${\mathbb Z}^2$-lattice and associated union of lattices, notation used in the paper}
\label{SETS}
\begin{figure}[h]
\begin{minipage}[h]{0.33\textwidth}
\begin{tikzpicture}
\draw[dashed] (-0.5,0)--(2.5,0);
\draw[dashed] (-0.5,1)--(2.5,1);
\draw[dashed] (-0.5,2)--(2.5,2);
\draw[dashed] (2,-0.5)--(2,2.5);
\draw[dashed] (1,-0.5)--(1,2.5);
\draw[dashed] (0,-0.5)--(0,2.5);
\filldraw
(0,0) circle (2pt) (1,0) circle (2pt) (2,0) circle (2pt) (0,1) circle (2pt) (0,2) circle (2pt) (1,2) circle (2pt) (1,1) circle (2pt) (2,1) circle (2pt) (2,2) circle (2pt);
\end{tikzpicture}
\end{minipage}
\begin{minipage}[h]{0.33\textwidth}
\begin{tikzpicture}
\draw[dashed] (-0.5,0)--(2.5,0);
\draw[dashed] (-0.5,1)--(2.5,1);
\draw[dashed] (-0.5,2)--(2.5,2);
\draw[dashed] (2,-0.5)--(2,2.5);
\draw[dashed] (1,-0.5)--(1,2.5);
\draw[dashed] (0,-0.5)--(0,2.5);
\draw[ultra thick] (-0.5,0)--(-0.1,0);
\draw[ultra thick] (0.1,0)--(0.9,0);
\draw[ultra thick] (1.1,0)--(1.9,0);
\draw[ultra thick] (2.1,0)--(2.5,0);
\draw[ultra thick] (-0.5,1)--(-0.1,1);
\draw[ultra thick] (0.1,1)--(0.9,1);
\draw[ultra thick] (1.1,1)--(1.9,1);
\draw[ultra thick] (2.1,1)--(2.5,1);
\draw[ultra thick] (-0.5,2)--(-0.1,2);
\draw[ultra thick] (0.1,2)--(0.9,2);
\draw[ultra thick] (1.1,2)--(1.9,2);
\draw[ultra thick] (2.1,2)--(2.5,2);
\end{tikzpicture}
\end{minipage}
\begin{minipage}[h]{0.33\textwidth}
\begin{tikzpicture}
\draw[dashed] (-0.5,0)--(2.5,0);
\draw[dashed] (-0.5,1)--(2.5,1);
\draw[dashed] (-0.5,2)--(2.5,2);
\draw[dashed] (2,-0.5)--(2,2.5);
\draw[dashed] (1,-0.5)--(1,2.5);
\draw[dashed] (0,-0.5)--(0,2.5);
\draw[ultra thick] (0,-0.5)--(0,-0.1);
\draw[ultra thick] (0,0.1)--(0,0.9);
\draw[ultra thick] (0,1.1)--(0,1.9);
\draw[ultra thick] (0,2.1)--(0,2.5);
\draw[ultra thick] (1,-0.5)--(1,-0.1);
\draw[ultra thick] (1,0.1)--(1,0.9);
\draw[ultra thick] (1,1.1)--(1,1.9);
\draw[ultra thick] (1,2.1)--(1,2.5);
\draw[ultra thick] (2,-0.5)--(2,-0.1);
\draw[ultra thick] (2,0.1)--(2,0.9);
\draw[ultra thick] (2,1.1)--(2,1.9);
\draw[ultra thick] (2,2.1)--(2,2.5);
\end{tikzpicture}
\end{minipage}
\newline

\bigskip

\begin{minipage}{0.33\textwidth}
\begin{tikzpicture}
\draw[dashed] (-0.5,0)--(2.5,0);
\draw[dashed] (-0.5,1)--(2.5,1);
\draw[dashed] (-0.5,2)--(2.5,2);
\draw[dashed] (2,-0.5)--(2,2.5);
\draw[dashed] (1,-0.5)--(1,2.5);
\draw[dashed] (0,-0.5)--(0,2.5);
\draw[ultra thick] (-0.5,-0.5)--(-0.1,-0.1);
\draw[ultra thick] (0.1,0.1)--(0.9,0.9);
\draw[ultra thick] (1.1,1.1)--(1.9,1.9);
\draw[ultra thick] (2.1,2.1)--(2.5,2.5);

\draw[ultra thick] (0.5,-0.5)--(0.9,-0.1);
\draw[ultra thick] (1.1,0.1)--(1.9,0.9);
\draw[ultra thick] (2.1,1.1)--(2.5,1.5);

\draw[ultra thick] (-0.5,0.5)--(-0.1,0.9);
\draw[ultra thick] (0.1,1.1)--(0.9,1.9);
\draw[ultra thick] (1.1,2.1)--(1.5,2.5);

\draw[ultra thick] (1.5,-0.5)--(1.9,-0.1);
\draw[ultra thick] (2.1,0.1)--(2.5,0.5);

\draw[ultra thick] (-0.5,1.5)--(-0.1,1.9);
\draw[ultra thick] (0.1,2.1)--(0.5,2.5);

\end{tikzpicture}
\end{minipage}
\begin{minipage}{0.33\textwidth}
\begin{tikzpicture}
\draw[dashed] (-0.5,0)--(2.5,0);
\draw[dashed] (-0.5,1)--(2.5,1);
\draw[dashed] (-0.5,2)--(2.5,2);
\draw[dashed] (2,-0.5)--(2,2.5);
\draw[dashed] (1,-0.5)--(1,2.5);
\draw[dashed] (0,-0.5)--(0,2.5);
\draw[ultra thick] (0.1,0.9)--(0.9,0.1);
\draw[ultra thick] (0.1,1.9)--(0.9,1.1);
\draw[ultra thick] (-0.1,0.1)--(-0.5,0.5);
\draw[ultra thick] (-0.1,1.1)--(-0.5,1.5);
\draw[ultra thick] (-0.1,2.1)--(-0.5,2.5);
\draw[ultra thick] (0.9,2.1)--(0.5,2.5);
\draw[ultra thick] (1.9,2.1)--(1.5,2.5);
\draw[ultra thick] (2.5,1.5)--(2.1,1.9);
\draw[ultra thick] (2.5,0.5)--(2.1,0.9);
\draw[ultra thick] (2.5,-0.5)--(2.1,-0.1);
\draw[ultra thick] (1.5,-0.5)--(1.1,-0.1);
\draw[ultra thick] (0.5,-0.5)--(0.1,-0.1);

\draw[ultra thick] (1.1,1.9)--(1.9,1.1);
\draw[ultra thick] (1.1,0.9)--(1.9,0.1);

\end{tikzpicture}
\end{minipage}
\begin{minipage}{0.33\textwidth}
\begin{tikzpicture}
\draw[dashed] (-0.5,0)--(2.5,0);
\draw[dashed] (-0.5,1)--(2.5,1);
\draw[dashed] (-0.5,2)--(2.5,2);
\draw[dashed] (2,-0.5)--(2,2.5);
\draw[dashed] (1,-0.5)--(1,2.5);
\draw[dashed] (0,-0.5)--(0,2.5);
\draw[fill] (0.1,0.1) rectangle (0.9,0.9);
\draw[fill] (0.1,1.1) rectangle (0.9,1.9);
\draw[fill] (1.1,0.1) rectangle (1.9,0.9);
\draw[fill] (1.1,1.1) rectangle (1.9,1.9);
\draw[fill] (-0.1,-0.1) rectangle (-0.5,-0.5);

\draw[fill] (-0.1,0.9) rectangle (-0.5,0.1);
\draw[fill] (-0.1,1.9) rectangle (-0.5,1.1);
\draw[fill] (-0.1,2.5) rectangle (-0.5,2.1);

\draw[fill] (0.9,2.5) rectangle (0.1,2.1);
\draw[fill] (1.9,2.5) rectangle (1.1,2.1);
\draw[fill] (2.5,2.5) rectangle (2.1,2.1);

\draw[fill] (2.5,1.9) rectangle (2.1,1.1);
\draw[fill] (2.5,0.9) rectangle (2.1,0.1);
\draw[fill] (2.5,-0.5) rectangle (2.1,-0.1);

\draw[fill] (1.9,-0.5) rectangle (1.1,-0.1);
\draw[fill] (0.9,-0.5) rectangle (0.1,-0.1);
\end{tikzpicture}
\end{minipage}

\caption{Sets associated with the ${\mathbb Z}^2-$lattice}
\label{sets}
\end{figure}
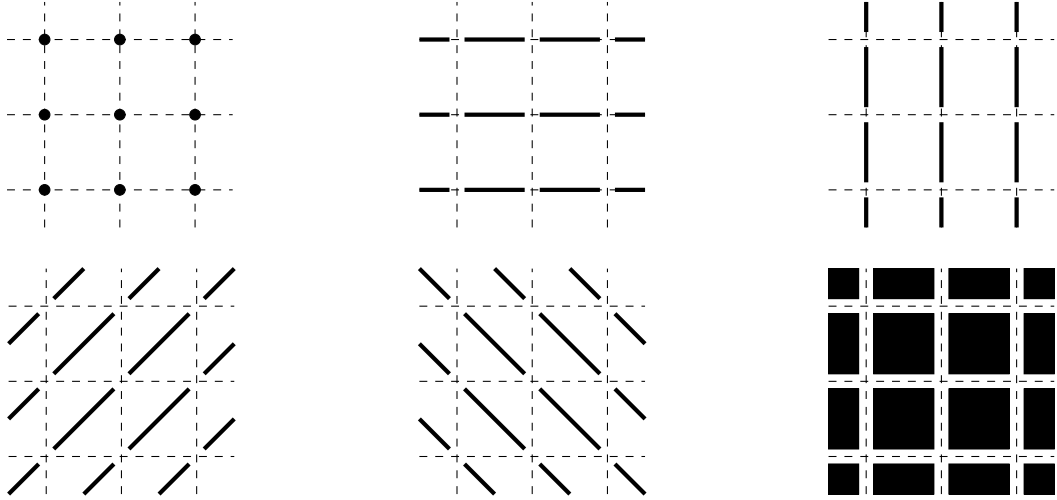

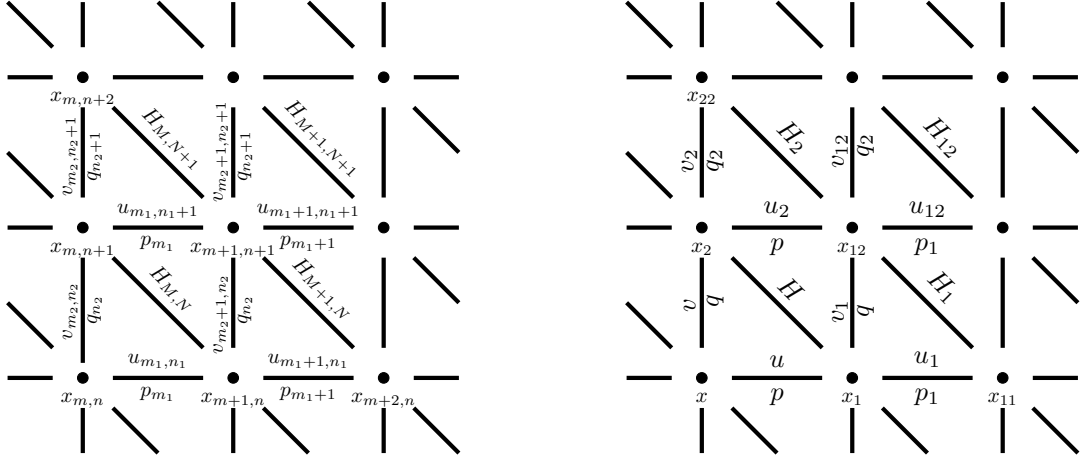
\begin{figure}[h]
\begin{minipage}[h]{0.5\textwidth}
\begin{tikzpicture}[scale=2, every node/.style={transform shape}];
\draw [fill] (0,0) circle [radius=1pt]   (0,-0.05) node [scale=0.4, below ] { $ x_{m,n}$};
\draw [fill] (1,0) circle [radius=1pt]   (1,-0.05) node [scale=0.4, below ] { $x_{m+1,n}$};
\draw [fill] (2,0) circle [radius=1pt]   (2,-0.05) node [scale=0.4, below]  { $x_{m+2,n}$};
\draw [fill] (0,1) circle [radius=1pt]   (0,0.95) node [scale=0.4,below ] {$x_{m,n+1}$};
\draw [fill] (0,2) circle [radius=1pt]   (0,1.95) node [scale=0.4,below ] {  $x_{m,n+2}$};
\draw [fill] (1,1) circle [radius=1pt]   (1,0.95) node [scale=0.4,below ] { $x_{m+1,n+1}$};
\draw [fill] (1,2) circle [radius=1pt];
\draw [fill] (2,1) circle [radius=1pt];
\draw [fill] (2,2) circle [radius=1pt];
\draw[ultra thick] (-0.5,0)--(-0.2,0)  (0.5,0) node [scale=0.4, above ] { $ u_{m_1,n_1}$}     (0.5,0) node [scale=0.4, below ] { $ p_{m_1}$};
\draw[ultra thick] (0.2,0)--(0.8,0);
\draw[ultra thick] (1.2,0)--(1.8,0)    (1.5,0) node [scale=0.4, above ] { $ u_{m_1+1,n_1}$} (1.5,0) node [scale=0.4, below ] { $ p_{m_1+1}$};
\draw[ultra thick] (2.2,0)--(2.5,0);
\draw[ultra thick] (-0.5,1)--(-0.2,1)  (0.5,1) node [scale=0.4, above ] { $ u_{m_1,n_1+1}$} (0.5,1) node [scale=0.4, below ] { $ p_{m_1}$};
\draw[ultra thick] (0.2,1)--(0.8,1);
\draw[ultra thick] (1.2,1)--(1.8,1)    (1.5,1) node [scale=0.4, above ] { $ u_{m_1+1,n_1+1}$}  (1.5,1) node [scale=0.4, below ] { $ p_{m_1+1}$};;
\draw[ultra thick] (2.2,1)--(2.5,1);
\draw[ultra thick] (-0.5,2)--(-0.2,2);
\draw[ultra thick] (0.2,2)--(0.8,2);
\draw[ultra thick] (1.2,2)--(1.8,2);
\draw[ultra thick] (2.2,2)--(2.5,2);
\draw[ultra thick] (0,-0.5)--(0,-0.2)   (-0.08,0.7) node [scale=0.4, left, rotate=90 ] {$v_{m_2,n_2}$}  (0.92,0.75) node [scale=0.4, left, rotate=90 ] {$v_{m_2+1,n_2}$} (0.08,0.3) node [scale=0.4, right, rotate=90 ] {$q_{n_2}$} (1.08,0.3) node [scale=0.4, right, rotate=90 ] {$q_{n_2}$} (0.08,1.25) node [scale=0.4, right, rotate=90 ] {$q_{n_2+1}$} (1.08,1.25) node [scale=0.4, right, rotate=90 ] {$q_{n_2+1}$};
\draw[ultra thick] (0,0.1)--(0,0.8)  (-0.08,1.8) node [scale=0.4, left, rotate=90 ] {$v_{m_2,n_2+1}$}  (0.92,1.9) node [scale=0.4, left, rotate=90 ] {$v_{m_2+1,n_2+1}$} ;
\draw[ultra thick] (0,1.2)--(0,1.8);
\draw[ultra thick] (0,2.2)--(0,2.5);
\draw[ultra thick] (1,-0.5)--(1,-0.2);
\draw[ultra thick] (1,0.2)--(1,0.8);
\draw[ultra thick] (1,1.2)--(1,1.8);
\draw[ultra thick] (1,2.2)--(1,2.5);
\draw[ultra thick] (2,-0.5)--(2,-0.2);
\draw[ultra thick] (2,0.2)--(2,0.8);
\draw[ultra thick] (2,1.2)--(2,1.8);
\draw[ultra thick] (2,2.2)--(2,2.5);
\draw[ultra thick] (0.2,0.8)--(0.8,0.2);
\draw[ultra thick] (0.2,1.8)--(0.8,1.2)     (0.5,0.5) node [scale=0.4, above, rotate=-45 ] {$H_{M,N}$} (1.5,0.5) node [scale=0.4, above, rotate=-45 ] {$H_{M+1,N}$} (0.5,1.5) node [scale=0.4, above, rotate=-45 ] {$H_{M,N+1}$} (1.5,1.5) node [scale=0.4, above, rotate=-45 ] {$H_{M+1,N+1}$};
\draw[ultra thick] (-0.2,0.2)--(-0.5,0.5);
\draw[ultra thick] (-0.2,1.2)--(-0.5,1.5);
\draw[ultra thick] (-0.2,2.2)--(-0.5,2.5);
\draw[ultra thick] (0.8,2.2)--(0.5,2.5);
\draw[ultra thick] (1.8,2.2)--(1.5,2.5);
\draw[ultra thick] (2.5,1.5)--(2.2,1.8);
\draw[ultra thick] (2.5,0.5)--(2.2,0.8);
\draw[ultra thick] (2.5,-0.5)--(2.2,-0.2);
\draw[ultra thick] (1.5,-0.5)--(1.2,-0.2);
\draw[ultra thick] (0.5,-0.5)--(0.2,-0.2);
\draw[ultra thick] (1.2,1.8)--(1.8,1.2);
\draw[ultra thick] (1.2,0.8)--(1.8,0.2);
\end{tikzpicture}
\end{minipage}
\begin{minipage}[h]{0.5\textwidth}
\begin{tikzpicture}[scale=2, every node/.style={transform shape}];
\draw [fill] (0,0) circle [radius=1pt]   (0,-0.05) node [scale=0.4, below ] { $ x$};
\draw [fill] (1,0) circle [radius=1pt]   (1,-0.05) node [scale=0.4, below ] { $x_1$};
\draw [fill] (2,0) circle [radius=1pt]   (2,-0.05) node[scale=0.4, below]  { $x_{11}$};
\draw [fill] (0,1) circle [radius=1pt]   (0,0.95) node [scale=0.4,below ] {$x_2$};
\draw [fill] (0,2) circle [radius=1pt]   (0,1.95) node [scale=0.4,below ] {  $x_{22}$};
\draw [fill] (1,1) circle [radius=1pt]   (1,0.95) node [scale=0.4,below ] { $x_{12}$};
\draw [fill] (1,2) circle [radius=1pt];
\draw [fill] (2,1) circle [radius=1pt];
\draw [fill] (2,2) circle [radius=1pt];
\draw[ultra thick] (-0.5,0)--(-0.2,0)  (0.5,0) node [scale=0.5, above ] { $ u$} node [scale=0.5, below ] { $ p$};
\draw[ultra thick] (0.2,0)--(0.8,0);
\draw[ultra thick] (1.2,0)--(1.8,0)    (1.5,0) node [scale=0.5, above ] { $ u_1$} node [scale=0.5, below ] { $ p_1$};
\draw[ultra thick] (2.2,0)--(2.5,0);
\draw[ultra thick] (-0.5,1)--(-0.2,1)  (0.5,1) node [scale=0.5, above ] { $ u_{2}$} node [scale=0.5, below ] { $ p$} ;
\draw[ultra thick] (0.2,1)--(0.8,1);
\draw[ultra thick] (1.2,1)--(1.8,1)    (1.5,1) node [scale=0.5, above ] { $ u_{12}$} node [scale=0.5, below ] { $ p_1$};
\draw[ultra thick] (2.2,1)--(2.5,1);
\draw[ultra thick] (-0.5,2)--(-0.2,2);
\draw[ultra thick] (0.2,2)--(0.8,2);
\draw[ultra thick] (1.2,2)--(1.8,2);
\draw[ultra thick] (2.2,2)--(2.5,2);
\draw[ultra thick] (0,-0.5)--(0,-0.2)   (-0.08,0.6) node [scale=0.5, left, rotate=90 ] {$v$} (0.09,0.4) node [scale=0.5, right, rotate=90 ] {$q$}  (0.92,0.6) node [scale=0.5, left, rotate=90 ] {$v_1$} (1.09,0.33) node [scale=0.5, right, rotate=90 ] {$q$} ;
\draw[ultra thick] (0,0.2)--(0,0.8)  (-0.08,1.6) node [scale=0.5, left, rotate=90 ] {$v_2$} (0.09,1.3) node [scale=0.5, right, rotate=90 ] {$q_2$}  (0.92,1.7) node [scale=0.5, left, rotate=90 ] {$v_{12}$} (1.09,1.4) node [scale=0.5, right, rotate=90 ] {$q_2$};
\draw[ultra thick] (0,1.2)--(0,1.8);
\draw[ultra thick] (0,2.2)--(0,2.5);
\draw[ultra thick] (1,-0.5)--(1,-0.2);
\draw[ultra thick] (1,0.2)--(1,0.8);
\draw[ultra thick] (1,1.2)--(1,1.8);
\draw[ultra thick] (1,2.2)--(1,2.5);
\draw[ultra thick] (2,-0.5)--(2,-0.2);
\draw[ultra thick] (2,0.2)--(2,0.8);
\draw[ultra thick] (2,1.2)--(2,1.8);
\draw[ultra thick] (2,2.2)--(2,2.5);
\draw[ultra thick] (0.2,0.8)--(0.8,0.2);
\draw[ultra thick] (0.2,1.8)--(0.8,1.2)     (0.5,0.5) node [scale=0.5, above, rotate=-45 ] {$H$} (1.5,0.5) node [scale=0.5, above, rotate=-45 ] {$H_1$} (0.5,1.5) node [scale=0.5, above, rotate=-45 ] {$H_2$} (1.5,1.5) node [scale=0.5, above, rotate=-45 ] {$H_{12}$};
\draw[ultra thick] (-0.2,0.2)--(-0.5,0.5);
\draw[ultra thick] (-0.2,1.2)--(-0.5,1.5);
\draw[ultra thick] (-0.2,2.2)--(-0.5,2.5);
\draw[ultra thick] (0.8,2.2)--(0.5,2.5);
\draw[ultra thick] (1.8,2.2)--(1.5,2.5);
\draw[ultra thick] (2.5,1.5)--(2.2,1.8);
\draw[ultra thick] (2.5,0.5)--(2.2,0.8);
\draw[ultra thick] (2.5,-0.5)--(2.2,-0.2);
\draw[ultra thick] (1.5,-0.5)--(1.2,-0.2);
\draw[ultra thick] (0.5,-0.5)--(0.2,-0.2);
\draw[ultra thick] (1.2,1.8)--(1.8,1.2);
\draw[ultra thick] (1.2,0.8)--(1.8,0.2);
\end{tikzpicture}
\end{minipage}
\caption{Fields on vertices, edges and diagonals of the ${\mathbb Z}^2$ lattice. \newline Standard notation (left figure) and concise notation used in the paper (right figure)}\label{notation}
\end{figure}

With the ${\mathbb Z}^2$ lattice 
one can associate (see Figure \ref{sets})
\begin{itemize}
\item the set of vertices $V=\left\{(m,n) \, | \,  m,n \in {\mathbb Z} \right\},$  which is the original ${\mathbb Z}^2$ lattice
\item the set of horizontal edges $E_h= \left\{ \{(m,n),(m+1,n)\} \, | \,  m,n \in {\mathbb Z} \right\} $
\item the set of vertical edges  $E_v= \left\{ \{(m,n),(m,n+1)\} \, | \,  m,n \in {\mathbb Z} \right\}$
\item the set of diagonals $D_{\diagup}= \left\{ \{(m,n),(m+1,n+1)\} \, | \,  m,n \in {\mathbb Z} \right\}$
\item the set of anti-diagonals  $D_{\, \diagdown}= \left\{ \{(m+1,n),(m,n+1)\} \, | \,  m,n \in {\mathbb Z} \right\}$
\item the set of faces $F= \left\{ \{(m,n),(m+1,n),(m,n+1),(m+1,n+1)\} \, | \,  m,n \in {\mathbb Z} \right\}$
\end{itemize}
  We label with $(m,n) \in {\mathbb Z}^2$,
$(m_1,n_1) \in {\mathbb Z}^2$, $(m_2,n_2) \in {\mathbb Z}^2$ and $(M,N) \in {\mathbb Z}^2,$ the elements of $V$, $E_h$, $E_v$ and $D_{\, \diagdown}$ respectively. Therefore, each of this sets can be regarded as a ${\mathbb Z}^2$ lattice itself.  In this way we also identify the sets $D_{\diagup}$ and $D_{\, \diagdown}$ with the set $F$.
On each of these sets, the same forward difference  operators ${\bf T_1},$  ${\bf T_2}$
are consistently defined, so, e.g. ${\bf T_1}: G_{M,N}\mapsto G_{M+1,N}$ and ${\bf T_2}: G_{M,N}
\mapsto G_{M,N+1}$. In what follows we use a  concise notation. We omit the independent variables and we denote the action of the shift operators ${\bf T_i},$ as
\[  {\bf T_i} G_{M,N}:=G_i, \quad {\bf T_i} {\bf T_i} G_{M,N}:=G_{ii}, \quad {\bf T_i} {\bf T_j}  G_{M,N}:=G_{ij}, \quad  i,j=1,2 \]
hence we are denoting  the forward shift in $i$-th direction by the subscript $i$ (see Figure \ref{notation}).
 Generalization of the considerations above to the ${\mathbb Z}^3$ lattice is straightforward. We have in addition a shift operator in the third direction   ${\bf T_3}: G_{M,N,K}\mapsto G_{M,N,K+1}:=G_3.$ We can  associate with the ${\mathbb Z}^3$ lattice a set of horizontal faces $F_h$, two sets  of vertical faces $F_{v_i}, i=1,2$ and the set of the centers of the cubes
\begin{itemize}
\item  the set of horizontal faces $F_h= \left\{ \{(m,n,l),(m+1,n,l),(m,n+1,l),(m+1,n+1,l)\} \, | \,  m,n,l \in {\mathbb Z} \right\}$
\item  two sets of vertical faces $F_{v_1}= \left\{ \{(m,n,l),(m+1,n,l),(m,n+1,l),(m+1,n+1,l)\} \, | \,  m,n,l \in {\mathbb Z} \right\},$
$F_{v_2}=\left\{ \{(m,n,l),(m+1,n,l),(m,n,l+1),(m+1,n,l+1)\} \, | \,  m,n,l \in {\mathbb Z} \right\}.$
\item the set of the centers of the cubes
$$
\begin{array}{c}
F_c=\left\{ \{(m,n,l),(m+1,n,l),(m,n+1,l),(m+1,n+1,l),(m,n,l+1),(m+1,n,l+1),\right.\\ [2mm]
 \left.(m,n+1,l+1),(m+1,n+1,l+1)\} \, | \,  m,n,l \in {\mathbb Z} \right\}.
\end{array}
$$
\end{itemize}
 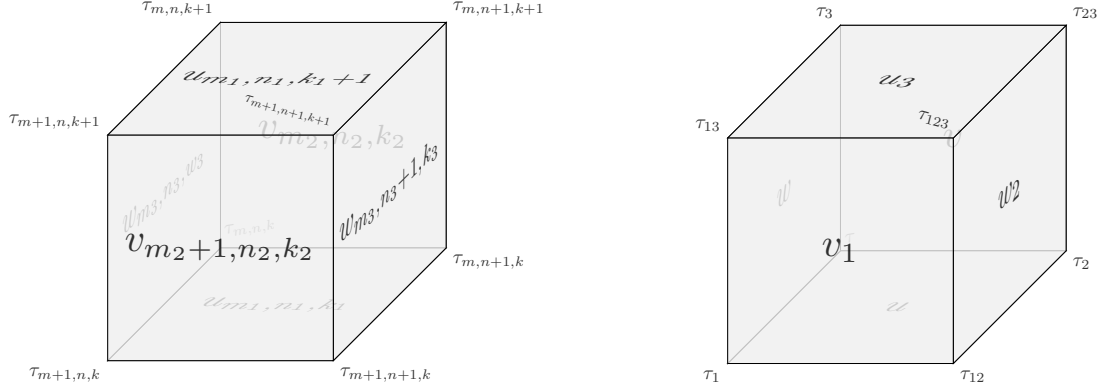
\begin{figure}[h]

 \begin{minipage}[b]{0.3\textwidth}
 \begin{tikzpicture}
    [x={(-0.5cm,-0.5cm)}, y={(1cm,0cm)}, z={(0cm,1cm)},
    scale=1.5,fill opacity=0.80,very thin,every node/.append style={transform shape}]
\newcommand\drawface{\draw[fill=gray!10] (-1,-1) rectangle (1,1)}

        \begin{scope}[canvas is yz plane at x=-1]
            \drawface;
            \node   {$v_{m_2,n_2,k_2}$};
        \end{scope}
        \begin{scope}[canvas is yx plane at z=-1]
           \drawface;
           \node[yscale=-1] {$u_{m_1,n_1,k_1}$};
        \end{scope}

        \begin{scope}[canvas is zx plane at y=-1]
           \drawface;
           \node[rotate=-90] {$w_{m_3,n_3,w_3}$};
        \end{scope}

        \begin{scope}[canvas is zx plane at y=1]
           \drawface;
           \node[rotate=-90] {$w_{m_3,n_3+1,k_3}$};
        \end{scope}
        \begin{scope}[canvas is yx plane at z=1]
          \drawface;
      \node[yscale=-1] {$u_{m_1,n_1,k_1+1}$};
      \end{scope}
      \begin{scope}[canvas is yz plane at x=1]
        \drawface;
        \node  {$v_{m_2+1,n_2,k_2}$};
      \end{scope}

 \draw    (-1,-1,-1) node [scale=0.5, above right, gray!30, rotate=10 ] { $ \tau_{m,n,k}$};
 \draw    (1,-1,-1) node [scale=0.5, below left,  rotate=0 ] { $ \tau_{m+1,n,k}$};
\draw    (1,-1,1) node [scale=0.5, above left,  rotate=0 ] { $ \tau_{m+1,n,k+1}$};
\draw    (1,1,-1) node [scale=0.5, below right,  rotate=0 ] { $ \tau_{m+1,n+1,k}$};
\draw    (-1,1,-1) node [scale=0.5, below right,  rotate=0 ] { $ \tau_{m,n+1,k}$};
\draw    (-1,-1,1) node [scale=0.5, above left,  rotate=0 ] { $ \tau_{m,n,k+1}$};
\draw    (-1,1,1) node [scale=0.5, above right,  rotate=0 ] { $ \tau_{m,n+1,k+1}$};
\draw    (1,1,1) node [scale=0.4, above left,  rotate=-15 ] { $ \tau_{m+1,n+1,k+1}$};
\end{tikzpicture}
\end{minipage} \hspace{4cm}
\begin{minipage}[b]{0.3\textwidth}
 \begin{tikzpicture}
    [x={(-0.5cm,-0.5cm)}, y={(1cm,0cm)}, z={(0cm,1cm)},
    scale=1.5,fill opacity=0.80,very thin,every node/.append style={transform shape}]
\newcommand\drawface{\draw[fill=gray!10] (-1,-1) rectangle (1,1)}

        \begin{scope}[canvas is yz plane at x=-1]
            \drawface;
            \node   {$v$};
        \end{scope}
        \begin{scope}[canvas is yx plane at z=-1]
           \drawface;
           \node[yscale=-1] {$u$};
        \end{scope}

        \begin{scope}[canvas is zx plane at y=-1]
           \drawface;
           \node[rotate=-90] {$w$};
        \end{scope}

        \begin{scope}[canvas is zx plane at y=1]
           \drawface;
           \node[rotate=-90] {$w_2$};
        \end{scope}
        \begin{scope}[canvas is yx plane at z=1]
          \drawface;
      \node[yscale=-1] {$u_3$};
      \end{scope}
      \begin{scope}[canvas is yz plane at x=1]
        \drawface;
        \node  {$v_1$};
      \end{scope}

 \draw    (-1,-1,-1) node [scale=0.5, above right, gray!30, rotate=10 ] { $ \tau$};
 \draw    (1,-1,-1) node [scale=0.5, below left,  rotate=0 ] { $ \tau_1$};
\draw    (1,-1,1) node [scale=0.5, above left,  rotate=0 ] { $ \tau_{13}$};
\draw    (1,1,-1) node [scale=0.5, below right,  rotate=0 ] { $ \tau_{12}$};
\draw    (-1,1,-1) node [scale=0.5, below right,  rotate=0 ] { $ \tau_{2}$};
\draw    (-1,-1,1) node [scale=0.5, above left,  rotate=0 ] { $ \tau_{3}$};
\draw    (-1,1,1) node [scale=0.5, above right,  rotate=0 ] { $ \tau_{23}$};
\draw    (1,1,1) node [scale=0.5, above left,  rotate=-15 ] { $ \tau_{123}$};
\end{tikzpicture}
\end{minipage}
\caption{Fields on vertices and faces of the ${\mathbb Z}^3$ lattice. \newline Standard notation (left figure) and concise notation used in the paper (right figure).}\label{notation3d}
\end{figure}

\section{Difference systems in bond and face variables}
\label{secb}

We are considering the following systems of difference equations, where $u_{m_1,n_1}$ and $v_{m_2,n_2}$ are the dependent variables, while $p_{m_1}$ and $q_{n_2}$ are prescribed functional parameters of one variable  (for notation see Figure \ref{notation} and \ref{notation3d}).

\begin{enumerate}
\item  The bond system associated to the (non-autonomous) Hirota's KdV equation \cite{KaNie1,KaNie3}
$$
\qquad u_2=v-p+q_2+(p_1-q)\frac{v}{u}, \quad v_1=u+p_1-q-(p-q_2)\frac{u}{v}, \hskip \textwidth minus  \textwidth   (H)
$$
its autonomous case, where $p$ and $q$  are regarded as constants
$$
\qquad u_2=v+(p-q)\left(-1+\frac{v}{u}\right), \quad v_1=u+(p-q)\left(1-\frac{u}{v}\right). \hskip \textwidth minus  \textwidth (H_a)
$$
Let us rewrite the $(H_a)$ system in a slightly different form and let us perform the following limiting procedure.
Replacing $(p,q)$ with $(\tilde p^2,\tilde q^2)$ and  making the substitutions $u=\tilde u (\tilde p- \tilde q) (\tilde p_1- \tilde q)$,\\
$v=\tilde v (\tilde p- \tilde q) (\tilde p- \tilde q_),$ the bond system $(H)$ reads
$$
\qquad \begin{array}{l}
{\displaystyle (\tilde p_1-\tilde q_2)\tilde u_2-(\tilde p-\tilde q) \tilde v=(\tilde p_1+\tilde q)\frac{\tilde v}{\tilde u}-(\tilde p+\tilde q_2)},\\ [3mm] {\displaystyle (\tilde p_1-\tilde q_2)\tilde v_1-(\tilde p-\tilde q) \tilde u=(\tilde p_1+\tilde q)-(\tilde p+\tilde q_2)\frac{\tilde u}{\tilde v}}.
 \end{array}\hskip \textwidth minus  \textwidth  (\tilde H)
$$
Replacing $\tilde u$ with  $\epsilon {\tilde {\tilde u}}+1$ and $ \tilde v$ with $\epsilon {\tilde {\tilde v}}+1,$  taking the limit $\epsilon\rightarrow 0,$    and removing  the tildes from   the resulting equations, we get:
$$
 (p_1-q_2)u_2=(p-q)v+(p_1+q) (v-u),\quad (p_1-q_2)v_1=(p-q)u+(p+q_2) (v-u). \quad (L)
$$
which in  autonomous case reads
$$
\qquad u_2={\dis v + \frac{p+q}{p-q}(v-u)},                     \quad v_1={\dis u + \frac{p+q}{p-q}(v-u)}. \hskip \textwidth minus  \textwidth (L_a)
$$

\item   The bond systems associated with some consistent-around-the-cube lattice equations \cite{Tasos}
$$
\begin{array}{lll}
u_2={\dis p v \, \frac{(1-q)u+q-p+(p-1)v}{q(1-p) u+p(q-1)v+(p-q)uv}},
                                                  & v_1={\dis q u \, \frac{(1-q)u+q-p+(p-1)v}{q(1-p) u+p(q-1)v+(p-q)uv}}
                                                                                                       & (F_{I})         \\ [3mm]
u_2={\dis \frac{v}{p} \,  \frac{pu-qv+q-p}{u-v}},     & v_1={\dis \frac{u}{q} \,  \frac{pu-qv+q-p}{u-v}}        & (F_{II})        \\ [3mm]
u_2={\dis \frac{v}{p} \,  \frac{pu-qv}{u-v}},     & v_1={\dis \frac{u}{q} \,  \frac{pu-qv}{u-v}}                 & (F_{III})       \\ [3mm]
u_2={\dis v\left(1+\frac{p-q}{u-v}\right)},         & v_1={\dis u \left(1+\frac{p-q}{u-v}\right)},         & (F_{IV})        \\ [3mm]
u_2={\dis v + \frac{p-q}{u-v}},                     & v_1={\dis u + \frac{p-q}{u-v}},                      & (F_V)        \\ [3mm]
\end{array}
$$
We recall that $p$ is a given function of the independent variable $m_1$ and $q$ is another given function of the variable $n_2,$ so  the equations above are in general non-autonomous.

\item  A system on faces of the ${\mathbb Z}^3$ lattice associated with the Hirota-Miwa equation \cite{Korepanov-1998,Sergeev-1998}
$$
\qquad u_3=\frac{uv}{u+w}, \quad  v_1=u+w, \quad w_2= \frac{vw}{u+w}. \hskip \textwidth minus  \textwidth (HM)
$$
\end{enumerate}

For all the difference systems presented here except for the $(HM)$, one can   associate the maps
$
\mathbb{CP}^1\times \mathbb{CP}^1\ni (u,v)\mapsto (U,V)=(u_2(u,v),v_1(u,v)) \in \mathbb{CP}^1\times \mathbb{CP}^1.
$ Many of these maps have been studied by various researchers  in connection with the Yang-Baxter property. The maps we get from  the bond systems $F_I-F_{V},$ are the F-list of quadrirational Yang-Baxter maps \cite{ABSf}. From $(L_a)$ we obtain a linear   Yang-Baxter map \cite{Franks-book,Kouloukas-2017,Dimakis2017}. From the bond system $(H)$ we obtain the (non-autonomous)  Hirota's KdV map \cite{KaNie1,KaNie3}, however this map is not Yang-Baxter in the strict sense but it satisfies an entwining Yang-Baxter relation \cite{koulou-entwin}. As a side remark, note that  the singularities of the map associated with the bond system $(\tilde H),$ under the limiting procedure to the system  $(L_a)$ described in $(i),$  merge into a single singularity~\cite{KaNie3}. In fact the Yang-Baxter maps associated with  $F_{II}-F_{V}$ difference systems, arise under certain limiting procedures that merge the singularities of the  map associated with the $F_I$ system \cite{ABSf}.     When the systems are defined on faces of a ${\mathbb Z}^3$ lattice, the associated maps reads $
\mathbb{CP}^1\times \mathbb{CP}^1\times \mathbb{CP}^1\ni(u,v,w)\mapsto (U,V,W)=(u_3(u,v,w),v_1(u,v,w),w_2(u,v,w)) \in \mathbb{CP}^1\times \mathbb{CP}^1\times \mathbb{CP}^1.
$
From the face system $(HM),$ we get  a functional tetrahedron map \cite{Korepanov-1998,Sergeev-1998}.

 An important  feature of the maps associated with $(H_a), (L_a), (F_I-F_V), (HM)$ is their involutivity. Involutive maps admit vast number of invariants and among them one can find invariants
with separated variables.  Following the procedure described in \cite{KaNie,KaNie1,KaNie3},  for each of the maps (except for $(HM)$), we present the  exhaustive list of invariants and alternating invariants with separated variables (see Table \ref{inttable}).

\begin{table}[h]
\caption{Exhaustive list of    invariants  with separated variables and alternating invariants with separated variables for the maps associated with the difference systems  \hbox{$(F_I-F_V)$}, $(H_a)$ and $(L_a) $ } \label{inttable}
\begin{tabular}{l|lll}
\toprule
Map &  Invariants &  &\\ [3mm]\hline
$F_{I}$
      & $\sqrt{\frac{q}{p}}\frac{U}{V}=\sqrt{\frac{p}{q}} \frac{ v}{u}$
      & $ \sqrt{\frac{q-1}{p-1}}\frac{U-1}{V-1}=\sqrt{\frac{p-1}{q-1}} \frac{v-1}{u-1}$
      & $\sqrt{\frac{q(q-1)}{p(p-1)}}\frac{U-p}{V-q}=\sqrt{\frac{p(p-1)}{q(q-1)}} \frac{v-q}{u-p}$\\ [3mm]
$F_{II}$
      & $\sqrt{\frac{p}{q}} \frac{U}{V}=\sqrt{\frac{q}{p}} \frac{v}{u}$
      & $\sqrt{\frac{p}{q}} \frac{U-1}{V-1}= \sqrt{\frac{q}{p}}\frac{v-1}{u-1}$
      & {\scriptsize $pU-\frac{1}{2}p-qV+\frac{1}{2}q=-(p u-\frac{1}{2}p-q v+\frac{1}{2}q) $}\\ [3mm]
$F_{III}$
      &  $\sqrt{\frac{p}{q}}\frac{U}{V}=\sqrt{\frac{q}{p}}\frac{v}{u}$
      & {\scriptsize  $pU-qV=-(pu-qv)$}
      &  $\frac{1}{U}-\frac{1}{V}=-(\frac{1}{u}-\frac{1}{v})$\\ [3mm]
$F_{IV}$
      &  $\frac{U}{V}=\frac{v}{u}$
      &  \parbox[t]{4cm}{\scriptsize $U+\frac{1}{2}p-V-\frac{1}{2} q=$ \\ \hbox{~~~ ~~~ ~~~} $-(u+\frac{1}{2}p-v-\frac{1}{2} q)$ }
      &  \parbox[t]{5.5cm}{\scriptsize $U^2+2pU+\frac{p^2}{2}-V^2-2qV-\frac{q^2}{2}=$  \\ \hbox{~~~ ~~~ ~~~} $ -(u^2+2pu+\frac{p^2}{2}-v^2-2qv-\frac{q^2}{2})$}\\ [3mm]
$F_{V}$
      &   {\scriptsize  $U-V=-(u-v)$}
      &   \parbox[t]{4cm}{\scriptsize $U^2+p-V^2-q=$ \\ \hbox{~~~ ~~~ ~~~} $-(u^2+p-v^2-q)$}
      &   \parbox[t]{5.5cm}{\scriptsize $U^3+3pU-V^3-3qV=$ \\ \hbox{~~~ ~~~ ~~~} $-(u^3+3pu-v^3-3qv)$}\\ [3mm]
$H_a$
      & {\scriptsize  $\frac{U}{V}=\frac{v}{u} $}
      &  {\scriptsize $(U+k)(V-k)=(u+k)(v-k)$}
      &    $\frac{V(U+k)}{U(V-k)}=\frac{v(u+k)}{u(v-k)}$\\ [3mm]
$L_a$
      & {\scriptsize  $U-V=-(u-v)$}
      &  {\scriptsize $\frac{U^2}{p}-\frac{V^2}{q}=\frac{u^2}{p}-\frac{v^2}{q}$}
      &  {\scriptsize {}}\\ [3mm]\hline

\bottomrule
\end{tabular}
\end{table}

 The existence of the invariants and/or the alternating invariants with separated variables, enable us to introduce a {\it potentialisation} procedure. This procedure is the ``opposite" to the vectorisation one that was explained in the introduction. It enable us to introduce a multi-parametric family of potentials, and we recover the lattice equations or the lattice relations (correspondences) out of the bond systems \cite{KaNie,KaNie1,KaNie3}. We illustrate how to apply the potentialisation procedure  to the bond system $(L_a)$ and to the face system $(HM),$  for the remaining cases we are referring  to \cite{KaNie,KaNie1,KaNie3}.

The invariants  and/or the the alternating  invariants with separated variables of the maps, on the level of the difference systems are responsible for the existence of conservation relations. These conservation relations guarantee the existence of a family of potentials. For example, lets consider the difference system~$(L_a).$
The associated map has  an alternating invariant $h_1(u,v)=u-v$  and the invariant $h_2(u,v)=\frac{u^2}{p}-\frac{v^2}{q}.$
The following conservation relations hold:
$$
u_2-v_1=v-u,\quad \frac{u_2^2}{p}-\frac{v_1^2}{q}=\frac{u^2}{p}-\frac{v^2}{q}.
$$
The first of the conservation relations above, guarantees the existence of a potential $f$ s.t. $u=f_1+f,$ $v=f_2+f$. In terms of the potential $f$ the difference system $(L_a)$ reads:
\begin{equation} \label{la1}
(p-q)(f_{12}-f)=(p+q)(f_2-f_1),
\end{equation}
which is a linear consistent-around-the-cube equation and serves as a special case of the classification results of \cite{Atkinson2009-linear} (see also \cite{Dimakis2017}).

The second   of the conservation relations above, guarantees the existence of a potential $g$ s.t. $\frac{u^2}{p}=g_i-g,$  $\frac{v^2}{q}=g_j-g$. In terms of the potential $g$ the difference system $(L_a)$ reads:
\begin{equation} \label{la2}
\begin{array}{l}
{\displaystyle [g_{12}-g_2+(g-g_1)k^2]^2 p^2+2(g-g_2)[g_{12}-g_2-(g-g_1)k^2](k+1)^2pq+} \\ [3mm]
\quad \qquad {\displaystyle (g-g_2)^2(k+1)^4q^2=0,}
\end{array}
\end{equation}
where ${\displaystyle k:=\frac{p+q}{p-q}}.$ The discriminant of this multiquadratic relation with respect to $g_{12}$ reads
 $$
 64  (p - q)^4  (p + q)^2  pq (g - g_1) (g - g_2),
 $$
 compare with \cite{AtkNie}.
Finally,  note that from  the defining relations of the potentials $f,g,$ by eliminating $u$ and $v$ we obtain:
$$
\frac{(f_1+f)^2}{p}=g_1-g, \quad \frac{(f_2+f)^2}{q}=g_2-g,
$$
which provides a B\"acklund transformation between these two equations. Moreover, by a linear combination of the potentials $f,g,$ a two-parameter family of potentials $h$ can be obtained
$$
h_1-h=a(-1)^{m+n}u+b\frac{u^2}{p},\quad h_2-h=a(-1)^{m+n}v+b\frac{v^2}{q}.
$$
In terms of the potential $h$ the difference system $(L_a)$ reads
$$
2a^2pq(p^2-q^2)A(h,h_1,h_2,h_{12})+b B(h,h_1,h_2,h_{12})=0,
$$
where the expressions $A(h,h_1,h_2,h_{12})$ and $B(h,h_1,h_2,h_{12}),$ are the left-hand-sides of the equations (\ref{la1}) and (\ref{la2}) expressed in terms of $h$ instead of $f$ and $g$ respectively.

\begin{table}[h]
\caption{ Some of the invariants in separated variables for the map associated with the difference system $(HM)$} \label{int-HM}
\begin{tabular}{l|llll}
Map & Invariants &  &\\ [3mm]\hline
$HM$
      & $UV=uv$
      & $ VW=vw$
      & $\frac{U}{W}=\frac{u}{w} $
       & $ U+V+W=u+v+w$\\ [3mm]
\bottomrule
\end{tabular}
\end{table}
Now we provide an example of how the potentialisation procedure works for the face system~$(HM)$. This procedure is basically the procedure described in \cite{doliwa-santini} as the introduction of the {\it first} and {\it second potentials}.     By using the first two invariants of Table \ref{int-HM}, we obtain  the following conservation relations:
$$
u_3 v_1=uv, \quad v_1w_2=vw.
$$
The first of the conditions above guarantees the existence of a function $f$ s.t. ${\displaystyle u=\frac{f_1}{f}},$   ${\displaystyle v=\frac{f}{f_3}},$  whereas the second condition guarantees the existence of a function $g$ s.t. ${\displaystyle v=\frac{g}{g_2}},$  ${\displaystyle w=\frac{g_1}{g}}.$
Following the terminology of the article \cite{doliwa-santini}, we refer the functions $f, g$  as first potentials. We observe that $f_3g=fg_2$ which guarantees the existence of a function $\tau$ (which we refer to as second potential) s.t. ${\displaystyle f=\frac{\tau}{\tau_2}},$ ${\displaystyle g=\frac{\tau}{\tau_3}}$. Expressing $u,v,w$ in terms of the $\tau-$function, we obtain:
$$
u=\frac{\tau_1\tau_2}{\tau \tau_{12}},\quad v=\frac{\tau \tau_{23}}{\tau_2 \tau_{3}},\quad w=\frac{\tau_1\tau_3}{\tau \tau_{13}}.
$$
 Finally, in terms of the $\tau-$function, the $(HM)$ difference system reads:
 \be \label{H-M-0}
 \tau  \tau_{123}- \tau_2  \tau_{13}- \tau_3  \tau_{12}=0,
  \ee
  which is the Hirota-Miwa integrable lattice equation in disguise.
Indeed by acting on (\ref{H-M-0}) with ${\bf T_{-2}T_{-3}},$ the later takes a form of  the Hirota-Miwa  equation.

\section{Bond systems in terms of their alternating invariants}\label{sec4}
In this section we rewrite the bond systems  in terms of some of their (alternating) invariants. We present here  only the  difference schemes  that
are proper recurrences. By proper recurrences we mean such recurrences  that there exists initial value problem that assures existence and uniqueness of the solution.
We arrive in two cases to well known quad equations,
and in four cases to pairs of   compatible 6-point equations. As we shall see, the pairs of the 6-point equations yield in turn
7-point equations.

\subsection{Difference integrable equations on quads and their potential versions, consistent-around-the-cube equations}
Before we proceed to the main subject of this section let us remind the Hirota discrete sine-Gordon equation
which we are going to refer here to as lattice potential sine-Gordon equation.
Table~\ref{disguise}  shows this equation in several disguises. Although all the equations are equivalent up to point transformations
they bear different names. We take the stand that it does not make sense to distinguish the particular forms of the equation.
Therefore, any of equations from the Table~\ref{disguise}  we refer to as lattice potential (Hirota's) sine-Gordon equation.

\begin{table}[h]
\caption{Left column: Disguises of lattice potential (Hirota's)  sine-Gordon  equation. Right Column: Point transformations which link the equations }
\label{disguise}
\begin{tabular}{l|l}
\toprule
Hirota's discrete sine-Gordon  equation  \cite{Hir-sG} & \\[3mm]
\hspace{1cm} $  p \sin (\psi_{12} + \psi +\psi_{1} +\psi_{2})=q \sin (\psi_{12} + \psi -\psi_{1} -\psi_{2})$ & \\[3mm]
\hline
Nonlinear Superposition Principle for sine-Gordon   \cite{Bianchi-1892} & \\ [3mm]
\hspace{1cm} $ (p-q)\tan (\omega_{12}-\omega)= (p+q) \tan (\omega_{1}-\omega_{2})$ & $\omega=(-1)^n \psi $ \\[3mm]
\hline
lattice potential modified KdV  \cite{Nijhoff-1995} & \\[3mm]
\hspace{1cm} $p (ww_1 - w_2w_{12}) = q (ww_2 - w_1w_{12}) $ &$w=e^{2i(-1)^n \psi}$ \cite{Faddeev-1994} \\[3mm]
\hline
$H3^0$  \cite{ABS} & \\[3mm]
\hspace{1cm} $p (xx_1 + x_2x_{12}) = q (xx_2 + x_1x_{12}) $ & $x=i^{m+n}e^{2i(-1)^n \psi}$ \cite{Faddeev-1994} \\[3mm]
\hline
noname & \\[3mm]
\hspace{1cm} $p (y_{12}y_1y_2 y-1) = q (y_{12} y-y_1 y_2) $ & $y=e^{2i\psi}$ \\[3mm]
\hline
\bottomrule
\end{tabular}
\end{table}

We are ready now to present the only two examples  when the systems of $F$-list rewritten in terms of their invariants  are equations on a quad.\footnote{Note that the linear bond system $(L_a)$ expressed in terms of its invariant $H=u-v$ leads to the following linear quad equation
$$
H_{12}-H=\frac{p+q_2}{p-q_2}H_2-\frac{p_1+q}{p_1-q}H_1.
$$
}
Namely,
expressing the difference system  $(F_V)$  in terms of its invariant $H=u-v$ we obtain the lattice Hirota's KdV equation
\[H_2(H_1 H_{12}+q-p_1)=H_1 (H H_2+q_2-p).\]
Whereas by expressing the difference system  $F_{III}$ in terms of its invariant ${\displaystyle H={\sqrt \frac{p}{q}} \frac{u}{v}}$, we obtain the lattice  sine-Gordon equation \cite{Volkov-1992,Bobenko-1993}
\[ H_{12} (\sqrt p H_2-\sqrt q_2)(\sqrt q H_1-\sqrt p_1)=H (\sqrt p_1 H_1-\sqrt q)(\sqrt q_2 H_2-\sqrt p).\]
Both cases have been widely discussed in the literature \cite{Volkov-1992,Bobenko-1993,BoPi,Faddeev-1994,NijOht}. These equations are related to equations that are consistent-around-the-cube. The
first equation, by the substitution $H=u-v=x_1-x_2,$ is related to lattice potential KdV equation ($H1$). The second equation, by the substitution ${\displaystyle H={\sqrt \frac{p}{q}} \frac{u}{v}}=\frac{x_1}{x_2},$ is related to lattice potential sine-Gordon equation (and hence $H3^0$) see Table~\ref{kdvsine}. Traditionally the adjective {\em potential} is added because of the connections of the type $H=x_1-x_2$, or ${\displaystyle H=\frac{x_1}{x_2}}$. Moreover, since potential versions of equations are consistent-around-the-cube, and therefore more popular (more in use), the remaining
equations are sometimes referred to as {\em non-potential} versions of the underlying consistent-around-the-cube equations.
The point is  one can find three parameter families of non-potential versions of given consistent-around-the-cube equations.
Only a few of them are proper recurrences. Two examples of these proper recurrences  are the lattice Hirota's KdV  and the lattice sine-Gordon equations and we present the remaining ones in the next subsection.
\begin{table}[h]
\caption{Lattice Korteweg de Vries and sine-Gordon equations}
\label{kdvsine}
\begin{tabular}{l|l|l}
\toprule
 &  $H_{12}-H = (T_1-T_2) \frac{p-q}{H} $ & $\displaystyle
\frac{H_{12}}{H}= T_1 \left(\frac{\sqrt{p} H-\sqrt q}{\sqrt q H-\sqrt{p}}\right)/ T_2 \left(\frac{\sqrt p H-\sqrt{q}}{\sqrt{q} H-\sqrt p}\right) .
$  \\ [3mm]
& lattice Hirota's KdV equation \cite{hirota-0} &  lattice sine Gordon equation \cite{Volkov-1992,Bobenko-1993} \\ [3mm]
&  & or lattice modified KdV equation \\  [3mm]  \hline
 &$H=x_1-x_2$ & ${\displaystyle H= \frac{x_1}{x_2}}$ \\  [3mm] \hline
 & $(x_{12}-x)(x_1-x_2)=p-q$ & $\sqrt p (x_1 x+x_{12}x_2)+\sqrt q (x_2x+x_{12}x_1) =0$\\ [3mm]
 & lattice potential KdV equation \cite{NC,WaEs} &  lattice potential Hirota sine Gordon \cite{Hir-sG} \\ [3mm]
& & or lattice potential modified KdV equation \\ [3mm]
\hline
\bottomrule
\end{tabular}
\end{table}

\subsection{Pairs of six point equations as non-potential versions of consistent-around-the-cube equations and seven point schemes associated with them.}
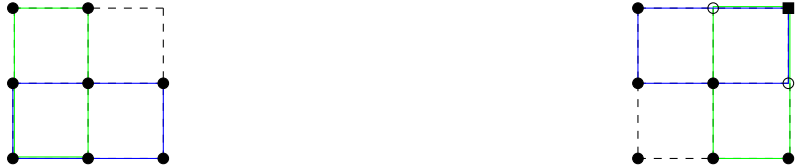
\begin{figure}[h] \centering
\begin{minipage}[c]{0.49\textwidth}\centering
\begin{tikzpicture}[scale=1]
\draw[dashed] (0,0)--(2,0);
\draw[dashed] (0,1)--(0,0);
\draw[dashed] (0.02,2)--(0.02,1);
\draw[dashed] (1,0)--(1,2);
\draw[dashed] (2,0)--(2,2);
\draw[dashed] (0,1)--(2,1);
\draw[dashed] (0,2)--(2,2);
\draw[green] (0.02,0)--(0.02,1)--(0.02,2)--(1,2)--(1,1)--(1,0.02)--(0,0.02);
\draw[blue] (0,0)--(1,0)--(2,0)--(2,1)--(1,1)--(0,1)--(0,0);
\filldraw
(0,0) circle (2pt) (1,0)circle (2pt) (2,0) circle (2pt) (0,1) circle (2pt) (0,2) circle (2pt) (1,2) circle (2pt) (1,1) circle (2pt) (2,1) circle (2pt);
\end{tikzpicture}
\end{minipage}
\hfill
\begin{minipage}[h]{0.49\textwidth}\centering
\begin{tikzpicture}[scale=1]
\draw[dashed] (0,0)--(2,0);
\draw[dashed] (0,1)--(0,0);
\draw[dashed] (0,2)--(0,1);
\draw[dashed] (1,0)--(1,2);
\draw[dashed] (2,1)--(2,2);
\draw[dashed] (2.02,1)--(2.02,0);
\draw[dashed] (0,1)--(2,1);
\draw[dashed] (0,2)--(2,2);
\draw[green] (1,0)--(2.02,0)--(2.02,1)--(2.02,2.02)--(1,2.02)--(1,1)--(1,0);
\draw[blue] (0,1)--(0,2)--(1,2)--(2,2)--(2,1)--(1,1)--(0,1);
\filldraw
(0,0) circle (2pt) (0,1) circle (2pt) (0,2) circle (2pt)  (1,0) circle (2pt) (2,0) circle (2pt) (1,1) circle (2pt) ;
\filldraw[black,yshift=-2pt,xshift=-2pt] (2,2) rectangle ++(4pt,4pt);
\draw[circle] (1,2) circle (2pt);\draw[circle] (2,1) circle (2pt);
\end{tikzpicture}
\end{minipage}
\caption{Left Figure: Points connected with blue line constitute the 6-point scheme for the $Q_E$ equation. Whereas, points connected with the green line constitute the 6-point scheme for the $Q_N$ equation.
 \newline Right Figure: Illustration of compatibility of the equations $Q_E$ and  $Q_N$. We prescribe initial data at the black points, then  the equations $Q_E$ and $Q_N$ provide unique values at open circles. The value of $H_{1122}$ (black rectangle) can be computed in two different ways, by using $T_2 Q_E$ or by $T_1 Q_N.$ Compatibility of the system of equations means  that these two ways of computing $H_{1122}$ give the same result.} \label{fig210}
\end{figure}

As for the remaining potential relations $H=f(x_i,x_j)$ that arise from Table \ref{inttable} we arrive at the systems of compatible six point equations (see Figure \ref{fig210})
$$
Q_E(H,H_1,H_2,H_{12},H_{11},H_{112})=0, \quad Q_N(H,H_1,H_2,H_{12},H_{22},H_{122})=0.
$$
To assure that these equations are proper recurrences, we confine ourselves  to the equations which have the following property:
\begin{itemize}
\item $Q_E$ is multilinear in variables $(H,H_2,H_{11},H_{112})$ and $Q_N$ is multilinear in  variables $(H,H_1,H_{22},H_{122}).$ We refer to this property as {\it multilinearity in the corners}.
    \end{itemize}
Indeed, prescribing the initial conditions on the two lines $\{(m,n)\in {\mathbb Z}^2 | m \in {\mathbb Z}, n=0\}$,
$\{(m,n)\in {\mathbb Z}^2 | m=0 , n\in {\mathbb Z} \}$ and at the point $(m,n)=(1,1)$ one can uniquely find the solution on the whole $ {\mathbb Z}^2$ lattice due to  multilinearity in the corners and due to compatibility of $Q^E$ and $Q^N$ equations (see Figure \ref{fig210}).

The prototypical example of such  equations is the system introduced by Nijhoff et.al  \cite{NijOht}
\[ (p+q)\tau_2 \tau_{11}+(p-q) \tau \tau_{112}=2p \tau_1\tau_{12},\quad  (p+q)\tau_1 \tau_{22}-(p-q) \tau \tau_{122}=2q \tau_2\tau_{12}, \]
which is a reformulation of the lattice Hirota's KdV equation in terms of the $\tau$ function.

We arrive at four examples
and we present them with the associated seven point equations i.e.,
first, the equation $Q_{EN}(H,H_1,H_2,H_{12},H_{112},H_{122},H_{1122})=0$
that arise from elimination of $H_{22}$ from the equations ${\bf T_2} Q_E(H,H_1,H_2,H_{12},H_{11},H_{112})=0$ and $Q_N(H,H_1,H_2,H_{12},H_{22},H_{122})=0$, second,  the equation $Q_{ES}(H_1,H_2,H_{11},H_{12},H_{22},H_{112},H_{122})=0$ that arise from elimination of $H$ from  the \\ $Q_E(H,H_1,H_2,H_{12},H_{11},H_{112})=0$ and $Q_N(H,H_1,H_2,H_{12},H_{22},H_{122})=0$
see Figures \ref{fig22} and \ref{fig33}. Now we list these systems.

\begin{figure}[h]
\begin{minipage}{0.5\textwidth}\centering
\begin{tikzpicture}[scale=1]
\draw[red,fill=red] (0,0) circle (2pt);
\draw[dashed] (0,0)--(2,0)--(2,2)--(0,2)--(0,0);
\draw[dashed] (1,0)--(1,2);

\draw[dashed] (0,1)--(2,1);
\draw[dashed] (0,2)--(2,2);
\draw (0,0)--(1,0)--(2,0)--(2,1)--(1,1)--(0,1)--(0,0);
\draw (1,0)--(1,1);
\draw (0,1)--(0,2)--(1,2)--(1,1);
\filldraw (1,0)circle (2pt) (2,0) circle (2pt) (0,1) circle (2pt) (0,2) circle (2pt) (1,2) circle (2pt) (1,1) circle (2pt) (2,1) circle (2pt);
\end{tikzpicture}
\end{minipage}\hfill
\begin{minipage}{0.5\textwidth}\centering
\begin{tikzpicture}[scale=1]
\draw[dashed] (0,0)--(2,0)--(2,2)--(0,2)--(0,0);
\draw[dashed] (1,0)--(1,2);

\draw[dashed] (0,1)--(2,1);
\draw[dashed] (0,2)--(2,2);
\draw[dashed] (0,0)--(1,0);

\draw (1,0)--(2,0)--(2,1)--(1,1)--(1,2)--(0,2)--(0,1)--(1,1)--(1,0);
\filldraw (1,0)circle (2pt) (2,0) circle (2pt) (0,1) circle (2pt) (0,2) circle (2pt) (1,2) circle (2pt) (1,1) circle (2pt) (2,1) circle (2pt);
\end{tikzpicture}
\end{minipage}
\caption{Elimination of the field $H$ from the schemes $Q_E$ and $Q_N$ (left Figure) leads to the 7-point scheme $Q_{ES}$ (right Figure) } \label{fig22}
\end{figure}
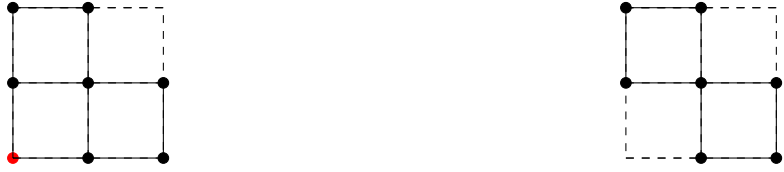


\begin{figure}[h]
\begin{minipage}{0.5\textwidth}\centering
\begin{tikzpicture}[scale=1]
\draw[red,fill=red] (0,2) circle (2pt);
\draw[dashed] (0,0)--(2,0)--(2,2)--(0,2)--(0,0);
\draw[dashed] (1,0)--(1,2);
\draw[dashed] (0,1)--(2,1);
\draw[dashed] (0,2)--(2,2);
\draw (0,0)--(1,0)--(1,1)--(0,1)--(0,0);\draw (0,1)--(0,2)--(1,2)--(2,2)--(2,1)--(1,1);
\draw (1,0)--(1,1);
\draw (0,1)--(0,2)--(1,2)--(1,1);
\filldraw
(0,0) circle (2pt) (1,0)circle (2pt) (2,2) circle (2pt) (0,1) circle (2pt)  (1,2) circle (2pt) (1,1) circle (2pt) (2,1) circle (2pt);
\end{tikzpicture}
\label{fig3-1}
\end{minipage}
\begin{minipage}{0.5\textwidth}\centering
\begin{tikzpicture}
\draw[dashed] (0,0)--(2,0)--(2,2)--(0,2)--(0,0);
\draw[dashed] (1,0)--(1,2);
\draw[dashed] (0,1)--(2,1);
\draw[dashed] (0,2)--(2,2);
\draw (0,0)--(1,0)--(1,1)--(0,1)--(0,0);\draw (1,1)--(2,1)--(2,2)--(1,2)--(1,1);
\filldraw
(0,0) circle (2pt) (1,0)circle (2pt) (2,2) circle (2pt) (0,1) circle (2pt)  (1,2) circle (2pt) (1,1) circle (2pt) (2,1) circle (2pt);
\end{tikzpicture}
\label{fig3-2}
\end{minipage}
\caption{Elimination of the field $H_{22}$ from the schemes ${\bf T_2} Q_E$ and $Q_N$ (left Figure) leads to the 7-point scheme $Q_{EN}$ (right Figure) } \label{fig33}
\end{figure}

$\bullet$   {\bf $F_{II}$ bond system in terms of the  invariant $H=\frac{u}{v}$ (pair of 6-point equations) and associated  7-point equations}

In case of the system $F_{II}$ we slightly change it by introducing the variables $u'=\sqrt{p}u$, $v'=\sqrt{q}v$. Omitting the primes the new form of the system reads
\[u_2 = v \frac{\sqrt{p} u - \sqrt{q} v + q - p}{\sqrt{q} u - \sqrt{p} v}, \qquad v_1 = u \frac{\sqrt{p} u - \sqrt{q} v + q - p}{\sqrt{q} u - \sqrt{p} v}.\]
Thanks to that one of the invariants  is  $H=\frac{u}{v}$. Rewriting the system in terms of $H$ we arrive at a pair of six point equations,
the first equation of type $Q_E$ reads
\begin{equation} \label{f2-1}
X_1 H_{12}  (k_1 A_2-k_2 A_1 H H_2)+ X {\bf T_1} (k_{2} B_{1} H_{2}H_{12}-k_{1}B_{2})= 0
\end{equation}
and the second one of type $Q_N$
\begin{equation} \label{f2-2}
 X_2 (k_1 A_2-k_2 A_1 H H_2 )+X H_2  {\bf T_2} (k_{2} B_{1} H_{2}H_{12}- k_{1} B_{2})  = 0
\end{equation}
where we introduced auxiliary variables
 \[ k:=p-q, \quad A:=\sqrt{p} H-\sqrt{q}, \quad B:=\sqrt{q}H-\sqrt{p},  \quad X:= H_{12} A_2 B_1-H A_1 B_2. \]
Out of these two six point equations we get a seven point equation of type
$Q_{ES}$
\begin{equation} \label{f2-3}
X_1   H_2 H_{12}  {\bf T_2} (k_{2} B_{1} H_{2}H_{12}- k_{1} B_{2})=
X_2 {\bf T_1} (k_{2} B_{1} H_{2}H_{12}-k_{1}B_{2})\\
\end{equation}
and a seven point equation of type
$Q_{EN}$
\begin{equation} \label{f2-4}
 k_{12} X_{12} X+ X_{12} A_{12}(k_1A_2-k_2A_1 H H_2) + X B_{12}  {\bf T_1 T_2} (k_2 B_1 H_2 H_{12}-k_1 B_2)=0
\end{equation}


We observe that $X=0$ solves each of these four equations (\ref{f2-1})-(\ref{f2-4}). Writing $X=0$ explicitly, namely  $H_{12} {\bf T_2} (\sqrt{p} H-\sqrt{q}) {\bf T_1} (\sqrt{q} H-\sqrt{p})-H {\bf T_1}(\sqrt{p} H-\sqrt{q}) {\bf T_2} (\sqrt{q} H-\sqrt{p})=0$ we recognize the lattice sine-Gordon equation. So every solution of lattice sine-Gordon equation is a solution of each of the four equations presented above.

$\bullet$ {\bf $F_{IV}$ bond system in terms of the  invariant $H=\frac{u}{v}$ (pair of 6-point equations) and associated  7-point equations}

If we rewrite the system $F_{IV}$ in terms of its invariant $H=\frac{u}{v}$
we get the
six point equation $Q_E$
\begin{equation} \label{QE-H}
\begin{array}{l}
 X_1 H_{12}  (k_1 G_2- k_2G_1H H_2)+
X {\bf T_1} (k_{2}G_{1}H_{2} H_{12} -k_{1}G_{2})=0,
\end{array}
\end{equation}
the six point equation $Q_N$
\begin{equation}
\begin{array}{l}
X_2 (k_1 G_2-k_2 G_1 H H_2 )+
X H_2  {\bf T_2} ( k_{2}G_{1}H_{2} H_{12}-k_{1}G_{2})=0,
\end{array}
\end{equation}
the seven point equation $Q_{ES}$
\begin{equation}
\begin{array}{l}
X_1 H_2 H_{12}  {\bf T_2} (k_{2}G_{1}H_{2} H_{12} -k_{1}G_{2})=
X_2  {\bf T_1} (k_{2}G_{1}H_{2} H_{12} -k_{1}G_{2}),
\end{array}
\end{equation}
the seven point equation $Q_{EN}$
\begin{equation}
\begin{array}{l}
\quad k_{12} X_{12} X+ X_{12} G_{12}(k_1G_2-k_2 G_1 H H_2) + X G_{12}  {\bf T_1 T_2} (k_2 G_1 H_2 H_{12}-k_1 G_2)=0,
\end{array}
\end{equation}
where this time we used the following auxiliary variables
\[k:=p-q, \quad G:=H-1, \quad X=G_1 G_2 (H_{12}-H).\]
The  equations above can be obtained from equations (\ref{f2-1})-(\ref{f2-4}) by the coalescence $A=B=:G,$
 in addition $H=f_{m-n}$ is a solution of the  equations above.

$\bullet$ {\bf $F_{IV}$ bond system in terms of the  invariant $H= u - v + p/2 - q/2$ (pair of 6-point equations) and associated  7-point equations}

This set of six points and seven points equations arises from system $F_{IV}$ rewritten in invariant $H= u - v + p/2 - q/2$.
Namely, denoting
\[k:=p-q, \quad A:=2H+p-q, \quad B:=2H-p+q, \quad
k_1 A_2-k_2 A_1=k_1 B_2-k_2 B_1=:X\] 
we get
the six point equation $Q_E$
\begin{equation}
\begin{array}{l}
X B_{12}  {\bf T_1} (B_{1}B_{12}- A A_{1})+
X_1 A_1  (B_2B_{12}-A A_2)=0,
\end{array}
\end{equation}
the six point equation $Q_N$
\begin{equation}
\begin{array}{l}
X B_{12}  {\bf T_2} (B_{2}B_{12}- A A_{2})+
X_2 A_2  (B_1B_{12}-A A_1)=0,
\end{array}
\end{equation}
the seven point equation $Q_{ES}$
\begin{equation}
\begin{array}{l}
X X_1 {\bf T_2} (B_{2}B_{12}- A A_{2})-X X_2 {\bf T_1} (B_{1}B_{12}- A A_{1})=
X_1 X_2 (A_1B_2-A_2B_1),
\end{array}
\end{equation}
the seven point equation $Q_{EN}$
\begin{equation}
\begin{array}{l}
X B_{12} B_{122}  {\bf T_1 T_2}(B_{1}B_{12}- A A_{1}) =
X_{12} A_2  A_{12} (B_{1}B_{12}- A A_{1}).
\end{array}
\end{equation}
Also in this case $X=0$ yields a set of particular solutions of the equations, this time the solutions are
$H=(p-q) f_{m+n}.$

$\bullet$ {\bf Non-autonomous Hirota's bond system   $(H)$  in terms of the  invariant $H=u/v$ (pair of 6-point equations) and associated  7-point equations}

The last set of six point and seven points equations arises from system $(H)$ rewritten in its invariant $H=u/v$.
Namely, by denoting
\[k:=p-q, \quad G:=k_2H-k_1, \quad X:=H-H_{12}\]
we get
the six point equation $Q_E$
\begin{equation}
\begin{array}{l}
{\displaystyle XH_1{\bf T_1}\left(G_1-HH_1G_2\right)-X_1 H_{11}H_{12}(G_1-H_1H_{12}G_2)+} \\
\quad \qquad {\displaystyle XX_1H_1H_{11}{\bf T_2}(G_1+H_1G)=0},
\end{array}
\end{equation}
the six point equation $Q_N$
\begin{equation}
\begin{array}{l}
{\displaystyle X H_1H_2 {\bf T_2}\left(G_1-HH_1G_2\right)-X_2 H_{12}(G_1-H_1H_{12}G_2)+} \\ [3mm]
\quad \qquad {\displaystyle XX_2H_1H_{12}{\bf T_2}(G+HG_2)=0.}
\end{array}
\end{equation}
The seven point equation $Q_{ES}$
\begin{equation}
\begin{array}{l}
{\displaystyle X_1H_2 H_{11}{\bf T_2}\left(G_1-H H_1G_2\right)-X_2{\bf T_1}(G_1-H H_1G_2)-}\\ [3mm]
\quad \qquad {\displaystyle X_1X_2H_{11}{\bf T_2}(G_1-H H_1G_2)=0}
\end{array}
\end{equation}
and the seven point equation $Q_{EN}$
\begin{equation}
\begin{array}{l}
{\displaystyle XH_1H_{12}{\bf T_1}{\bf T_2}\left(G_1-H H_1G_2\right)-X_{12}H_{12}H_{112}(G_1-H_1H_{12}G_2)+} \\ [3mm]
\quad \qquad {\displaystyle X X_{12}H_1H_{112}{\bf T_2}\left(G_1+H_1G_{12}+H_1G\right)=0.}
\end{array}
\end{equation}
Also in this case $X=0$ yields to a special solution for all the equations presented above. This solution reads:
$H= f_{m-n}.$

\subsection{A non-potential form of the Hirota-Miwa equation}
We denote the three invariants (presented in Table \ref{int-HM}) of the $(HM)$ face system  as:
\be \label{hm-rv}
I=u v,\quad  J=v w, \quad H=\frac{u}{w}.
\ee
 Expressing the $(HM)$ system in terms of $I,J$ we get:
\be \label{n-poy-hm}
\frac{I_{13}I_3}{I_1I}=\frac{(I+J)_3}{(I+J)_1},\quad
 \frac{I_{23}I_3}{I_2I(I+J)_3}=\frac{J_{23}J_2}{J_3J(I+J)_2},\quad  \frac{J_{12}J_2}{J_1J}=\frac{(I+J)_2}{(I+J)_1}.
\ee
These equations can be rewritten in terms of the invariant $H=I/J$ only, namely
\begin{equation}\label{hm-pot}
H_{123} H (1+H_3)(1+H_{12})=H_{12}H_3 (1+H_{13}) (1+H_2).
\end{equation}
  Equation (\ref{hm-pot}) can be regarded as the non-potential version of the Hirota-Miwa equation. However, in the literature it is referred to as the gauge invariant form of the Hirota-Miwa equation \cite{Kuniba-1994,Krichever-1997,Doliwa-2000}.

\section{Elimination of Variables}\label{sec5}

In this section we rewrite each of the bond systems presented in the introduction, in terms of the $u$ variable
only by the elimination of the  dependent variable $v$ and its shift $v_1$.
From the bond systems $(F_I-F_V)$ we obtain multi-quadratic quad correspondences. Whereas for the bond system (L), we obtain a linear quad equation. For the bond system associated to the Hirota map we obtain a multilinear quad equation. Finally, for the difference system (HM) we obtain a system of compatible 10-point equations as well as 10-point and 12-point equations.

 As an illustrative example we show how the elimination procedure works in the case of the  $F_{III}$ bond system.
Namely, the $F_{III}$ bond system reads:
\be \label{F30}
u_2=\frac{v}{p}\frac{pu-q v}{u-v},\qquad v_1=\frac{u}{q}\frac{pu-q v}{u-v}.
\ee
Shifting the first equation of $(\ref{F30})$ in the  first direction we obtain
\be \label{F3i}
u_{12}=\frac{v_1}{p_1}\frac{p_1 u_1-q v_1}{u_1-v_1}.
\ee
From equations  $(\ref{F30})$  and  $(\ref{F3i})$ we  eliminate the fields $v, v_1$  and we obtain the equation  $d Q3^{0*}$ (see also Table \ref{QQ-systems})
$$
(y_1-y) (z_1 y-z y_1)-q (z_1-z)^2=0,\quad \mbox{where} \quad z=p u u_2,\;\; y=p(u+u_2).
$$

\subsection{A quad equation associated to the bond system of Hirota's map}

Applying the elimination procedure to the non-autonomous Hirota's difference system $(H)$
$$
u_2=v-k_2+k_1\frac{v}{u}, \quad v_1=u+k_1-k_2\frac{u}{v}, \quad  \mbox{where} \quad k:=p-q
$$
we obtain
\begin{equation} \label{t0}
\frac{(u_{12}+k_{12})}{(u_1+k_{11})} \frac{(u_{2}+k_{2})}{(u+k_{1})} =\frac{u_2}{u_1}.
\end{equation}
For the autonomous case, where $p,q$ are considered constant, the system  $(H)$ reduces to $(H_a)$ and from the equation above we obtain
\begin{equation} \label{t1}
\frac{u_1u_{12}-k^2}{uu_2-k^2}=\frac{u_1+k}{u_2+k}.
\end{equation}
Equations (\ref{t0}) and (\ref{t1}) are examples of  asymmetric multilinear quad equations. Similar equations have been presented in \cite{hyd-general,Adler_2011}. We have not been able to identify  yet the lattice equations (\ref{t0}) and (\ref{t1})  with  known ones.

Finally, when we eliminate the fields $v, v_1$  from the system $(L)$
we get the linear equation
$$
k (k_1+k_{11})(k+k_1-k_2)u-(k+k_1)k_{11}k_{12}u_1+(k_1+k_{11})k_2k_{12}u_2-(k+k_1)k_{12}k_{112}u_{12}=0.
$$

\subsection{Multi-quadratic quad correspondences associated to the bond systems of quadrirational Yang-Baxter maps}
In \cite{AtkNie}  a list of integrable multi-quadratic quad equations were introduced. Their construction is  based  on  the  hypothesis  that
discriminants  of  the  defining  polynomials  factorise  in  a  particular  way  that  allows
to  reformulate  the  equation  as  a  single-valued  system \cite{JJJ}. More precisely, a list of 8 multi-quadratic quad equation $Q1^*, Q2^*, Q3^*, Q4^*,A1^*, A2^*, H2^*, H3^*,$ were derived with the property that the discriminants of these polynomials   wrt. any of its arguments is proportional to the factor of {\em two}  bi-quadratic polynomials. These polynomials are the ones used by Adler et.al. \cite{ABS2} in their list of integrable quad equations. There exists a generalisation of the construction presented in \cite{AtkNie}. Namely,  in \cite{j-atk} it was introduced a multi-quadratic polynomial respecting the symmetry of the Fano plane, such that its discriminants  wrt. any of its arguments is proportional to the factor of {\em three} bi-quadratic polynomials.

 Here we show that from the bond systems $F_I-F_V$ one can obtain multi-quadratic quad equations which are degenerate cases of the $Q1^*, Q2^*, Q3^*, Q4^*$ and $Q3^{0*}$ (where the superscript "0" stands for $Q3^*$ with $\delta=0$) equations.  Namely, by applying the elimination procedure at the bond systems  $F_I-F_V,$ we obtain five  multi-quadratic quad equations which are presented in Table \ref{QQ-systems}. We denote these degenerate cases     as $d Q1^*, d Q2^*, d Q3^*, d Q4^*$ and $dQ3^{0*}$ respectively. The degeneration lies in the fact   that the discriminants of the polynomials   wrt. any of its arguments is proportional to  {\em one} bi-quadratic polynomial. The proportionality factor is a perfect square of an expression so in principle it can be removed by dividing the multiquadratic equation with this expression.  In Table \ref{Q0-systems} we present the discriminants of the obtained multiquadratic equations wrt. $u_{12}$ which are bi-quadratic polynomials of $u$ and $u_2$ (we omit the perfect square factors since they can be removed by  appropriate divisions of the original equations).


\begin{table}[h]
\caption{ Multi-quadratic quad equations obtained from bond systems $F_I-F_{V}$} \label{QQ-systems}
\begin{tabular}{l|l}
\toprule
Name & Equation  \\ [3mm]\hline
$dQ{4}^*$   & $(y_1-y)(z_1-z+z_1 y-z y_1)-q (z_1-z)(y_1-y+z_1y-zy_1)=0,$ \\ [3mm]
            &  $z:=\frac{(u-p)(u_2-p)}{p(p-1)}$, $y:=\frac{(u-1)(u_2-1)}{p-1}$ \\[3mm]
$dQ{3}^*$   &  $(y_1-y)(z_1y-zy_1)-q(z_1-z)^2+q(z_1-z)(y_1-y)=0,$       \\ [3mm]
            &  $z:=p u u_2$, $y:=p(-1+u+u_2) $ \\ [3mm]
$d Q3^{0*}$ &   $(y_1-y)(z_1y-zy_1)-q(z_1-z)^2=0,$ \\ [3mm]
            &  $z:=p u u_2$,  $y:=p (u+u_2)$\\ [3mm]
$dQ{2}^*$  & $(y_1-y)(y_1 z-y z_1)+(z_1-z)^2+q(z_1-z)(y_1-y)=0,$  \\ [3mm]
            &  $z:=u u_2$, $ y:=p+u+u_2$  \\ [3mm]
$dQ{1}^*$  &  $(y_1-y)(y_1 z-yz_1)-(z_1-z)^2-q (y_1-y)^2=0,$   \\ [3mm]
            &  $z:=p-uu_2$, $y:=u+u_2$\\ [3mm] \hline
\bottomrule
\end{tabular}
\end{table}

\begin{table}[h!]
\caption{Discriminants of multiquadratic quad equations of the Table $\ref{QQ-systems}$ (with omitted perfect square factors) } \label{Q0-systems}
\begin{tabular}{l}
\toprule
  Discriminant\\ [3mm]\hline
$\Delta(dQ4^*,u_{12})$  $\varpropto$        $(y-q z-q+1)^2-4q(q-1)z$\\ [3mm]
$\Delta(dQ3^*,u_{12})$ $\varpropto$   $(y+q)^2-4qz$\\ [3mm]
$\Delta(d Q3^{0*},u_{12})$ $\varpropto$  $y^2-4qz$\\ [3mm]
$\Delta(dQ2^*,u_{12})$ $\varpropto$   $(y-q)^2-4z$\\ [3mm]
$\Delta(dQ1^*,u_{12})$ $\varpropto$  $y^2+4(z-q)$\\ [3mm]\hline
\bottomrule
\end{tabular}
\end{table}

\subsection{10-point compatible equations on the $\mathbb{Z}^3$ lattice associated to the Hirota-Miwa difference system}
A difference system associated to the Hirota-Miwa equation reads:
\begin{eqnarray} \label{a}
u_3=\frac{uv}{u+w}, \\ \label{b}
v_1=u+w, \\ \label{c}
w_2=\frac{vw}{u+w}.
\end{eqnarray}
 In what follows, we apply the elimination procedure to the face system (\ref{a}),(\ref{b}),(\ref{c}).

\subsubsection{Elimination of the fields $v,w$}
Shifting (\ref{a}) by $\bf T_1$ and by using (\ref{b}) we eliminate $v_1$ to obtain:
\be \label{a1}
w_1=-u_1\left(1-\frac{u+w}{u_{13}}\right).
\ee
Shifting (\ref{c}) by  $\bf T_1$ and by using (\ref{b}) we eliminate $v_1,$ also we  shift (\ref{a1}) by $\bf T_2$ and we obtain the equations
$$
w_{21}=\frac{(u+w)w_1}{u_1+u+w}, \quad w_{12}=-u_{12}\left(1-\frac{u_2+w_2}{u_{132}}\right).
$$
From the last two equations since $(w_{12}=w_{21})$ we eliminate $w_{12}$ and from from the resulting equation by the use of (\ref{a1}) and (\ref{c}) we eliminate $w_1, w_2$  to obtain
\be \label{d}
v w u_{12}=(u+w)[u_{123}(u+u_{12}-u_{13}+w)-u_2u_{12}].
\ee
Now we are solving the system of  equations  (\ref{a}) and (\ref{d}) for  $v$ and $w$ and we get
\be \label{A}
v=u_3\frac{u_{123}(u_{12}-u_{13})+u_{12}(u_3-u_2)}{u_3u_{12}-u u_{123}}, \quad
w=u\frac{u_{123}(u-u_{13})+u_{12}(u_{123}-u_2)}{u_3u_{12}-u u_{123}}.
\ee
Finally, we use (\ref{A}) to eliminate the fields $v, w$ and their shifts from $(\ref{a})-(\ref{c}).$ Then, equation (\ref{a})  is identically satisfied, whereas equations $(\ref{b}), $(\ref{c}) leads to the following system of homogeneous quintic   $10-$point equations which are compatible on the 3-dimensional lattice.

\be
\begin{array}{l} \label{B1}
{\displaystyle  X u_{13}{\bf T}_1\left(u_{123}(u_{12}-u_{13})+u_{12}(u_3-u_2)\right)=X_1u\left(u_{123}(u_{12}-u_{13})+u_{12}(u_3-u_2)\right) },
\end{array}
\ee
\be
\begin{array}{l}
\label{B2}
{\displaystyle X u_2{\bf T}_2\left(u_{123}(u-u_{13})+u_{12}(u_{123}-u_2)\right)=X_2u_3\left(u_{123}(u-u_{13})+u_{12}(u_{123}-u_2)\right)},
\end{array}
\ee
where
$$
X:=uu_{123}-u_3u_{12}
$$
 Here compatibility means that the equations provide unique values at points $u_{1123}$ and $u_{1223}$ and the two different ways to evaluate $u_{11223}$ leads to the same result (see Figure \ref{comp}). A consequence of this compatibility is that by imposing a correct initial value problem f.i. the initial data  provided on the 3 perpendicular  coordinate planes,  the solution exists on the whole $\mathbb{Z}^3$ lattice.

Note that  by acting on (\ref{B1}) with ${\bf T_{-2} T_{-3}}$ followed by  the change of the independent variables $(m,n)\mapsto(-n,-m)$, the  later takes exactly the form of a 10-point equation firstly presented in \cite{Fu-Nijhoff-2017}, which in this paper it is referred to as {\it non-potential version of the Hirota-Miwa} equation.

\begin{center}
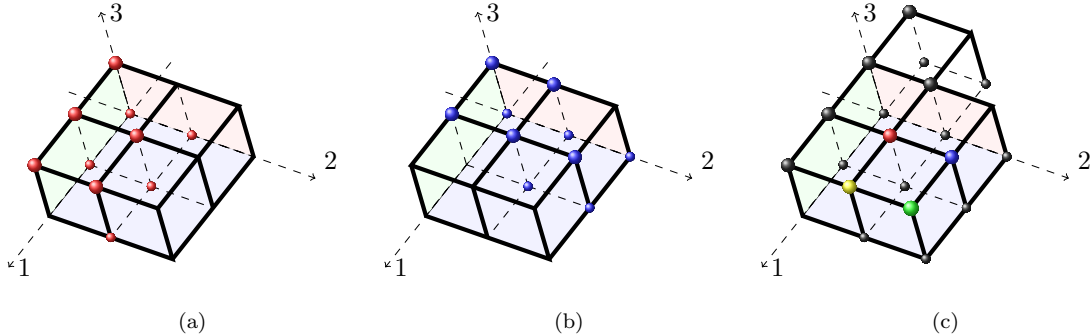
\begin{figure}[htb]
\begin{minipage}[h]{0.3\textwidth}
\begin{tikzpicture}[scale=1,tdplot_rotated_coords,
                    cube/.style={very thin,black},
                    grid/.style={very thin,gray},
                    axis/.style={->,blue,ultra thick},
                    rotated axis/.style={->,black,very thin,dashed}]

    \foreach \x in {0,...,2}
       \foreach \y in {0,...,2}
       \foreach \z in {0,...,1}
       {
           \draw[grid] (\x,\y,0) -- (\x,\y,1);
            \draw[grid] (0,\y,\z) -- (2,\y,\z);
                        \draw[grid] (\x,0,\z) -- (\x,2,\z);
        }

    \draw[cube,dashed,fill=blue!5] (0,0,0) -- (0,2,0) -- (2,2,0) -- (2,0,0) -- cycle;

    \draw[cube,dashed,fill=red!5] (0,0,0) -- (0,2,0) -- (0,2,1) -- (0,0,1) -- cycle;

    \draw[cube,dashed,fill=green!5] (0,0,0) -- (2,0,0) -- (2,0,1) -- (0,0,1) -- cycle;

    \draw[rotated axis] (-1,0,0) -- (3,0,0) node[anchor=west]{$1$};
    \draw[rotated axis] (0,-1,0) -- (0,3,0) node[anchor=south west]{$2$};
    \draw[rotated axis] (0,0,0) -- (0,0,2) node[anchor=west]{$3$};

 \foreach \x in {0,1,2}
   \foreach \y in {0,1,2}
      \foreach \z in {0,1}{
          \ifthenelse{  \lengthtest{\x pt < 2pt}  }
           {
              True
                \draw [black,dashed]   (\x,\y,\z) -- (\x+1,\y,\z);
           }
           { False
           }
         \ifthenelse{  \lengthtest{\y pt < 2pt}  }
           {
             \draw [black,dashed]   (\x,\y,\z) -- (\x,\y+1,\z);
           }
           {
           }
         \ifthenelse{  \lengthtest{\z pt < 1pt}  }
           {
                \draw [black,dashed]   (\x,\y,\z) -- (\x,\y,\z+1);
           }
           {
           }}

\draw [black,ultra thick]   (0,0,1)--(2,0,1)--(2,2,1)--(0,2,1)--(0,0,1);
\draw [black,ultra thick]   (2,0,1)--(2,0,0)--(2,2,0)--(2,2,1);
\draw [black,ultra thick]   (0,2,1)--(0,2,0)--(2,2,0);
\draw [black,ultra thick]   (2,1,0)--(2,1,1)--(0,1,1);
\draw [black,ultra thick]   (1,2,0)--(1,2,1)--(1,0,1);


\shade[rotated axis,ball color = red!80] (0,0,0) circle (0.06cm);
\shade[rotated axis,ball color = red!80] (1,0,0) circle (0.06cm);
\shade[rotated axis,ball color = red!80] (0,1,0) circle (0.06cm);
\shade[rotated axis,ball color = red!80] (0,0,1) circle (0.09cm);
\shade[rotated axis,ball color = red!80] (1,1,0) circle (0.06cm);
\shade[rotated axis,ball color = red!80] (1,0,1) circle (0.09cm);
\shade[rotated axis,ball color = red!80] (2,0,1) circle (0.09cm);
\shade[rotated axis,ball color = red!80] (1,1,1) circle (0.09cm);
\shade[rotated axis,ball color = red!80] (2,1,1) circle (0.09cm);
\shade[rotated axis,ball color = red!80] (2,1,0) circle (0.06cm);

\end{tikzpicture}
\captionsetup{font=footnotesize}
\captionof*{figure}{(a)}
\end{minipage}
\begin{minipage}[h]{0.3\textwidth}
\begin{tikzpicture}[scale=1,tdplot_rotated_coords,
                    cube/.style={very thin,black},
                    grid/.style={very thin,gray},
                    axis/.style={->,blue,ultra thick},
                    rotated axis/.style={->,black,very thin,dashed}]

    \foreach \x in {0,...,2}
       \foreach \y in {0,...,2}
       \foreach \z in {0,...,1}
       {
           \draw[grid] (\x,\y,0) -- (\x,\y,1);
            \draw[grid] (0,\y,\z) -- (2,\y,\z);
                        \draw[grid] (\x,0,\z) -- (\x,2,\z);
        }

    \draw[cube,dashed,fill=blue!5] (0,0,0) -- (0,2,0) -- (2,2,0) -- (2,0,0) -- cycle;

    \draw[cube,dashed,fill=red!5] (0,0,0) -- (0,2,0) -- (0,2,1) -- (0,0,1) -- cycle;

    \draw[cube,dashed,fill=green!5] (0,0,0) -- (2,0,0) -- (2,0,1) -- (0,0,1) -- cycle;

    \draw[rotated axis] (-1,0,0) -- (3,0,0) node[anchor=west]{$1$};
    \draw[rotated axis] (0,-1,0) -- (0,3,0) node[anchor=south west]{$2$};
    \draw[rotated axis] (0,0,0) -- (0,0,2) node[anchor=west]{$3$};

 \foreach \x in {0,1,2}
   \foreach \y in {0,1,2}
      \foreach \z in {0,1}{
          \ifthenelse{  \lengthtest{\x pt < 2pt}  }
           {
              True
                \draw [black,dashed]   (\x,\y,\z) -- (\x+1,\y,\z);
           }
           { False
           }
         \ifthenelse{  \lengthtest{\y pt < 2pt}  }
           {
             \draw [black,dashed]   (\x,\y,\z) -- (\x,\y+1,\z);
           }
           {
           }
         \ifthenelse{  \lengthtest{\z pt < 1pt}  }
           {
                \draw [black,dashed]   (\x,\y,\z) -- (\x,\y,\z+1);
           }
           {
           }}

\draw [black,ultra thick]   (0,0,1)--(2,0,1)--(2,2,1)--(0,2,1)--(0,0,1);
\draw [black,ultra thick]   (2,0,1)--(2,0,0)--(2,2,0)--(2,2,1);
\draw [black,ultra thick]   (0,2,1)--(0,2,0)--(2,2,0);
\draw [black,ultra thick]   (2,1,0)--(2,1,1)--(0,1,1);
\draw [black,ultra thick]   (1,2,0)--(1,2,1)--(1,0,1);


\shade[rotated axis,ball color = blue!80] (0,0,0) circle (0.06cm);
\shade[rotated axis,ball color = blue!80] (0,0,1) circle (0.09cm);
\shade[rotated axis,ball color = blue!80] (0,1,0) circle (0.06cm);
\shade[rotated axis,ball color = blue!80] (1,1,0) circle (0.06cm);
\shade[rotated axis,ball color = blue!80] (1,0,1) circle (0.09cm);
\shade[rotated axis,ball color = blue!80] (0,2,0) circle (0.06cm);
\shade[rotated axis,ball color = blue!80] (0,1,1) circle (0.09cm);
\shade[rotated axis,ball color = blue!80] (1,1,1) circle (0.09cm);
\shade[rotated axis,ball color = blue!80] (1,2,0) circle (0.06cm);
\shade[rotated axis,ball color = blue!80] (1,2,1) circle (0.09cm);

\end{tikzpicture}
\captionsetup{font=footnotesize}
\captionof*{figure}{(b)}
\end{minipage}
\begin{minipage}[h]{0.3\textwidth} 
\begin{tikzpicture}[scale=1,tdplot_rotated_coords,
                    cube/.style={very thin,black},
                    grid/.style={very thin,gray},
                    axis/.style={->,blue,ultra thick},
                    rotated axis/.style={->,black,very thin,dashed}]

    \foreach \x in {0,...,2}
       \foreach \y in {0,...,2}
       \foreach \z in {0,...,1}
       {
           \draw[grid] (\x,\y,0) -- (\x,\y,1);
            \draw[grid] (0,\y,\z) -- (2,\y,\z);
                        \draw[grid] (\x,0,\z) -- (\x,2,\z);
        }

    \draw[cube,dashed,fill=blue!5] (0,0,0) -- (0,2,0) -- (2,2,0) -- (2,0,0) -- cycle;

    \draw[cube,dashed,fill=red!5] (0,0,0) -- (0,2,0) -- (0,2,1) -- (0,0,1) -- cycle;

    \draw[cube,dashed,fill=green!5] (0,0,0) -- (2,0,0) -- (2,0,1) -- (0,0,1) -- cycle;

    \draw[rotated axis] (-1,0,0) -- (3,0,0) node[anchor=west]{$1$};
    \draw[rotated axis] (0,-1,0) -- (0,3,0) node[anchor=south west]{$2$};
    \draw[rotated axis] (0,0,0) -- (0,0,2) node[anchor=west]{$3$};

 \foreach \x in {0,1,2}
   \foreach \y in {0,1,2}
      \foreach \z in {0,1}{
          \ifthenelse{  \lengthtest{\x pt < 2pt}  }
           {
              True
                \draw [black,dashed]   (\x,\y,\z) -- (\x+1,\y,\z);
           }
           { False
           }
         \ifthenelse{  \lengthtest{\y pt < 2pt}  }
           {
             \draw [black,dashed]   (\x,\y,\z) -- (\x,\y+1,\z);
           }
           {
           }
         \ifthenelse{  \lengthtest{\z pt < 1pt}  }
           {
                \draw [black,dashed]   (\x,\y,\z) -- (\x,\y,\z+1);
           }
           {
           }}

\draw [black,ultra thick]   (0,0,1)--(2,0,1)--(2,2,1)--(0,2,1)--(0,0,1);
\draw [black,ultra thick]   (2,0,1)--(2,0,0)--(2,2,0)--(2,2,1);
\draw [black,ultra thick]   (0,2,1)--(0,2,0)--(2,2,0);
\draw [black,ultra thick]   (2,1,0)--(2,1,1)--(0,1,1);
\draw [black,ultra thick]   (1,2,0)--(1,2,1)--(1,0,1);

\draw [black,ultra thick]   (0,0,1)--(-1,0,1); \draw [black,ultra thick] (0,1,1)--(-1,1,1)--(-1,1,0); \draw [black,ultra thick]   (-1,0,1)--(-1,1,1);
\draw [black,dashed]   (-1,0,1)--(-1,0,0)--(-1,1,0)--(0,1,0);

\shade[rotated axis,ball color = black!80] (0,0,0) circle (0.06cm);
\shade[rotated axis,ball color = black!80] (0,0,1) circle (0.09cm);
\shade[rotated axis,ball color = black!80] (0,1,0) circle (0.06cm);
\shade[rotated axis,ball color = black!80] (1,1,0) circle (0.06cm);
\shade[rotated axis,ball color = black!80] (0,2,0) circle (0.06cm);
\shade[rotated axis,ball color = black!80] (0,1,1) circle (0.09cm);
\shade[rotated axis,ball color = black!80] (1,2,0) circle (0.06cm);
\shade[rotated axis,ball color = black!80] (1,1,1) circle (0.09cm);
\shade[rotated axis,ball color = black!80] (2,1,1) circle (0.09cm);
\shade[rotated axis,ball color = blue!80] (1,2,1) circle (0.09cm);
\shade[rotated axis,ball color = black!80] (2,2,0) circle (0.06cm);
\shade[rotated axis,ball color = black!80] (2,2,1) circle (0.09cm);

\shade[rotated axis,ball color = black!80] (-1,0,0) circle (0.06cm);
\shade[rotated axis,ball color = black!80] (-1,1,0) circle (0.06cm);
\shade[rotated axis,ball color = black!80] (-1,0,1) circle (0.09cm);
\shade[rotated axis,ball color = black!80] (1,0,0) circle (0.06cm);
\shade[rotated axis,ball color = black!80] (1,0,1) circle (0.09cm);
\shade[rotated axis,ball color = black!80] (1,1,0) circle (0.06cm);
\shade[rotated axis,ball color = black!80] (2,1,0) circle (0.06cm);
\shade[rotated axis,ball color = black!80] (2,0,1) circle (0.09cm);
\shade[rotated axis,ball color = black!80] (1,2,0) circle (0.06cm);
\shade[rotated axis,ball color = red!80] (1,1,1) circle (0.09cm);
\shade[rotated axis,ball color = yellow!80] (2,1,1) circle (0.09cm);
\shade[rotated axis,ball color = black!80] (2,2,0) circle (0.06cm);
\shade[rotated axis,ball color = green!80] (2,2,1) circle (0.1cm);

\end{tikzpicture}
\captionsetup{font=footnotesize}
\captionof*{figure}{(c)}
\end{minipage}
\caption{(a): The 10-point scheme for the equation (\ref{B1}). (b): The 10-point scheme for the equation~(\ref{B2}). (c): Illustration of compatibility of the equations (\ref{B1}) and  (\ref{B2}). We prescribe initial data at the black points. Equation (\ref{B1}) shifted by ${\bf T_{-1}}$ provides a unique value of $x_{123}$ (red point), next by using the equation (\ref{B1}) we obtain the value of $x_{1123}$ (yellow point). Then,  the equation (\ref{B2})  provides a unique value at the blue point.  The value of $x_{11223}$ (green point) can be computed in two different ways, by the equation (\ref{B1}) shifted by ${\bf T_2}$ or by (\ref{B2}) shifted by ${\bf T_1}$. Here these two ways of computing $x_{11223}$ give the same result, a manifestation of compatibility.}\label{comp}
\end{figure}
\end{center}

Now  by shifting the equation (\ref{B1}) by $\bf T_2$ and by using (\ref{B2}) to eliminate $u_{22}$  we arrive at the following homogeneous
 10-point quartic equation on the $\mathbb{Z}^3$ lattice (the analogue of the 7-point equation on the $\mathbb{Z}^2$ lattice). It reads:
\be \label{10pt-2}
X_{12}\left( u (u_3-u_2)- u_3(u_{13}-u_{12})\right)+
X {\bf T_{1} T_{2}} \left(u_{123} (u_{13}-u_{12})-u_{12}(u_{3}-u_{2})\right)=0
\ee
Whereas, from (\ref{B1}) and (\ref{B2}) by eliminating the variable $u,$  we obtain the following 12-point homogeneous sextic equation
\be \label{12pt-1}
\begin{array}{l}
{\displaystyle X_1 {\bf T_2}\left( u_{1}u_{12}\left(u(u_{2}-u_{123})+u_{3}(u_{13}-u)\right)\right)+}\\ [3mm]
\quad \qquad {\displaystyle X_2{\bf T_1}\left(u_{3} u_{23}\left(u_{123}(u-u_{13})+u_{12}(u_{123}-u_{2})\right)\right)=0}.
\end{array}
\ee
\begin{center}
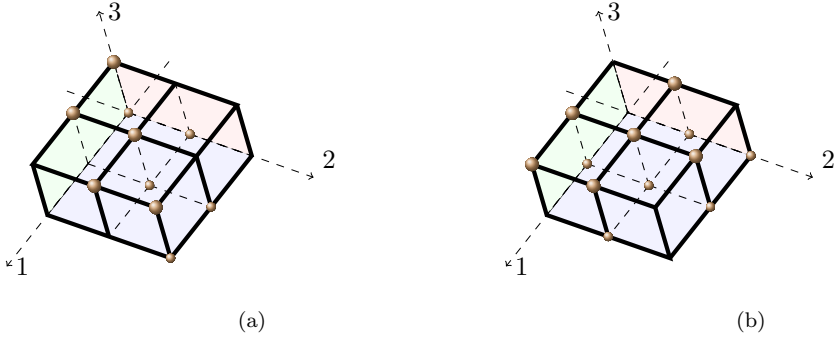
\begin{figure}[htb]
\centering
\begin{minipage}[b]{0.4\textwidth}
\begin{tikzpicture}[scale=1,tdplot_rotated_coords,
                    cube/.style={very thin,black},
                    grid/.style={very thin,gray},
                    axis/.style={->,blue,ultra thick},
                    rotated axis/.style={->,black,very thin,dashed}]

    \foreach \x in {0,...,2}
       \foreach \y in {0,...,2}
       \foreach \z in {0,...,1}
       {
           \draw[grid] (\x,\y,0) -- (\x,\y,1);
            \draw[grid] (0,\y,\z) -- (2,\y,\z);
                        \draw[grid] (\x,0,\z) -- (\x,2,\z);
        }

    \draw[cube,dashed,fill=blue!5] (0,0,0) -- (0,2,0) -- (2,2,0) -- (2,0,0) -- cycle;

    \draw[cube,dashed,fill=red!5] (0,0,0) -- (0,2,0) -- (0,2,1) -- (0,0,1) -- cycle;

    \draw[cube,dashed,fill=green!5] (0,0,0) -- (2,0,0) -- (2,0,1) -- (0,0,1) -- cycle;

    \draw[rotated axis] (-1,0,0) -- (3,0,0) node[anchor=west]{$1$};
    \draw[rotated axis] (0,-1,0) -- (0,3,0) node[anchor=south west]{$2$};
    \draw[rotated axis] (0,0,0) -- (0,0,2) node[anchor=west]{$3$};

 \foreach \x in {0,1,2}
   \foreach \y in {0,1,2}
      \foreach \z in {0,1}{
          \ifthenelse{  \lengthtest{\x pt < 2pt}  }
           {
              True
                \draw [black,dashed]   (\x,\y,\z) -- (\x+1,\y,\z);
           }
           { False
           }
         \ifthenelse{  \lengthtest{\y pt < 2pt}  }
           {
             \draw [black,dashed]   (\x,\y,\z) -- (\x,\y+1,\z);
           }
           {
           }
         \ifthenelse{  \lengthtest{\z pt < 1pt}  }
           {
                \draw [black,dashed]   (\x,\y,\z) -- (\x,\y,\z+1);
           }
           {
           }}

\draw [black,ultra thick]   (0,0,1)--(2,0,1)--(2,2,1)--(0,2,1)--(0,0,1);
\draw [black,ultra thick]   (2,0,1)--(2,0,0)--(2,2,0)--(2,2,1);
\draw [black,ultra thick]   (0,2,1)--(0,2,0)--(2,2,0);
\draw [black,ultra thick]   (2,1,0)--(2,1,1)--(0,1,1);
\draw [black,ultra thick]   (1,2,0)--(1,2,1)--(1,0,1);


\shade[rotated axis,ball color = brown!80] (0,0,0) circle (0.06cm);
\shade[rotated axis,ball color = brown!80] (0,1,0) circle (0.06cm);
\shade[rotated axis,ball color = brown!80] (1,1,0) circle (0.06cm);
\shade[rotated axis,ball color = brown!80] (1,2,0) circle (0.06cm);
\shade[rotated axis,ball color = brown!80] (2,2,0) circle (0.06cm);

\shade[rotated axis,ball color = brown!80] (0,0,1) circle (0.09cm);
\shade[rotated axis,ball color = brown!80] (1,0,1) circle (0.09cm);
\shade[rotated axis,ball color = brown!80] (1,1,1) circle (0.09cm);
\shade[rotated axis,ball color = brown!80] (2,1,1) circle (0.09cm);
\shade[rotated axis,ball color = brown!80] (2,2,1) circle (0.09cm);
\end{tikzpicture}
\captionsetup{font=footnotesize}
\captionof*{figure}{(a)}
\end{minipage}
\begin{minipage}[b]{0.4\textwidth}
\begin{tikzpicture}[scale=1,tdplot_rotated_coords,
                    cube/.style={very thin,black},
                    grid/.style={very thin,gray},
                    axis/.style={->,blue,ultra thick},
                    rotated axis/.style={->,black,very thin,dashed}]

    \foreach \x in {0,...,2}
       \foreach \y in {0,...,2}
       \foreach \z in {0,...,1}
       {
           \draw[grid] (\x,\y,0) -- (\x,\y,1);
            \draw[grid] (0,\y,\z) -- (2,\y,\z);
                        \draw[grid] (\x,0,\z) -- (\x,2,\z);
        }

    \draw[cube,dashed,fill=blue!5] (0,0,0) -- (0,2,0) -- (2,2,0) -- (2,0,0) -- cycle;

    \draw[cube,dashed,fill=red!5] (0,0,0) -- (0,2,0) -- (0,2,1) -- (0,0,1) -- cycle;

    \draw[cube,dashed,fill=green!5] (0,0,0) -- (2,0,0) -- (2,0,1) -- (0,0,1) -- cycle;

    \draw[rotated axis] (-1,0,0) -- (3,0,0) node[anchor=west]{$1$};
    \draw[rotated axis] (0,-1,0) -- (0,3,0) node[anchor=south west]{$2$};
    \draw[rotated axis] (0,0,0) -- (0,0,2) node[anchor=west]{$3$};

 \foreach \x in {0,1,2}
   \foreach \y in {0,1,2}
      \foreach \z in {0,1}{
          \ifthenelse{  \lengthtest{\x pt < 2pt}  }
           {
              True
                \draw [black,dashed]   (\x,\y,\z) -- (\x+1,\y,\z);
           }
           { False
           }
         \ifthenelse{  \lengthtest{\y pt < 2pt}  }
           {
             \draw [black,dashed]   (\x,\y,\z) -- (\x,\y+1,\z);
           }
           {
           }
         \ifthenelse{  \lengthtest{\z pt < 1pt}  }
           {
                \draw [black,dashed]   (\x,\y,\z) -- (\x,\y,\z+1);
           }
           {
           }}

\draw [black,ultra thick]   (0,0,1)--(2,0,1)--(2,2,1)--(0,2,1)--(0,0,1);
\draw [black,ultra thick]   (2,0,1)--(2,0,0)--(2,2,0)--(2,2,1);
\draw [black,ultra thick]   (0,2,1)--(0,2,0)--(2,2,0);
\draw [black,ultra thick]   (2,1,0)--(2,1,1)--(0,1,1);
\draw [black,ultra thick]   (1,2,0)--(1,2,1)--(1,0,1);


\shade[rotated axis,ball color = brown!80] (1,0,0) circle (0.06cm);
\shade[rotated axis,ball color = brown!80] (0,1,0) circle (0.06cm);
\shade[rotated axis,ball color = brown!80] (1,1,0) circle (0.06cm);
\shade[rotated axis,ball color = brown!80] (1,0,1) circle (0.09cm);
\shade[rotated axis,ball color = brown!80] (0,2,0) circle (0.06cm);
\shade[rotated axis,ball color = brown!80] (0,1,1) circle (0.09cm);

\shade[rotated axis,ball color = brown!80] (2,1,0) circle (0.06cm);
\shade[rotated axis,ball color = brown!80] (1,2,0) circle (0.06cm);
\shade[rotated axis,ball color = brown!80] (1,1,1) circle (0.09cm);
\shade[rotated axis,ball color = brown!80] (2,1,1) circle (0.09cm);
\shade[rotated axis,ball color = brown!80] (1,2,1) circle (0.09cm);
\shade[rotated axis,ball color = brown!80] (2,0,1) circle (0.09cm);
\end{tikzpicture}
\captionsetup{font=footnotesize}
\captionof*{figure}{(b)}
\end{minipage}
\caption{(a): The 10-point scheme for the equation~(\ref{10pt-2}). (b): The 12-point scheme for the equation~(\ref{12pt-1})}
\end{figure}
\end{center}
Note that $X=0$ serves as a special solution of (\ref{10pt-2}) and of (\ref{12pt-1}), as well as for the system (\ref{B1}) (\ref{B2}). Explicitly this solution reads
$$
u_{l,m,n}=f_{l-m,n} g_{l,m},
$$
where $f, g$ are arbitrary functions of $2$ independent variables.
\subsubsection{Elimination of the fields $u,w$}
Working similarly, we can express $u, w$ in terms of $v$ and its shifts to obtain:
\be \label{Y1}
u=\frac{v_1v_{13}}{v}\frac{v_{12}(v_{123}-v_3)+v_2(v-v_{12})}{v_2v_{13}-v_3v_{12}}, \quad w=\frac{v_1v_{12}}{v}\frac{v_{13}(v_2-v_{123})+v_3(v_{13}-v)}{v_2v_{13}-v_3v_{12}}.
\ee
The compatible system of $10-$points equations now reads:
\be \label{B11}
Xv_{133}{\bf T_3}\left(v_{12}(v_{123}-v_3)+v_2(v-v_{12}) \right)= X_3 v_3 \left(v_{12}(v_{123}-v_3)+v_2(v-v_{12}) \right),
\ee
\be \label{B22}
  Xv_{122} {\bf T_2} \left( v_{13}(v_{123}-v_2)+v_3(v-v_{13}) \right)=X_2v_2\left( v_{13}(v_{123}-v_2)+v_3(v-v_{13}) \right),
\ee
where we have introduced the variable
$$
X:=v_2v_{13}-v_3v_{12}.
$$
From (\ref{B11}) and (\ref{B22}) by eliminating $v,$  we obtain the following 10-point homogeneous quartic equation
\be \label{10pt-22}
X_2{\bf T_3} \left(v_{13}(v_{123}-v_{3})+v_{3}(v-v_{12})\right)+ X_3{\bf T_2}\left(v_{12}(v_{3}-v_{123})+v_{2}(v_{12}-v)\right)=0,
\ee
 On the other hand, if we shift the  equation  (\ref{B11}) by $\bf T_2$ and we eliminate $v_{22}$ by the use of  (\ref{B22}), we obtain:
\be \label{12pt-22}
X_{23}v_3v_{23}\left(v_{12}(v_{123}-v_2)+v_2(v-v_{13})\right)=X{\bf T_2 T_3} \left(v_{1}v_{12}\left(v_{13}(v_{123}-v_{2})+v_{3}(v-v_{13})\right)\right) ,
\ee
that is a 12-point homogeneous sextic equation. Here it holds that $X=0$ is a special solution for the system (\ref{B11}),(\ref{B22}), as well as for the equations (\ref{10pt-22}) and (\ref{12pt-22}). Explicitly this solution reads
$$
v_{l,m,n}=F_{l,m+n} G_{m,n},
$$
where $F, G$ are arbitrary functions of $2$ independent variables.

\section{Conclusions} \label{sec6}

The theory of 2-dimensional discrete integrable systems has focused so far mainly  on equations defined on quads. Equations defined on higher point stencils of a two dimensional lattice appeared occasionally, for instance the discrete analogue of the Toda-lattice that is defined on a five point stencil \cite{Hirota1977},  an equation defined on a hexagonal configuration \cite{Date1983}, the discrete analogue of the Boussisesq equation  \cite{Nijhoff1992-gelfand} that is defined on a nine point stencil and various  six point and seven point schemes  \cite{NijOht,Suris1996,Rizos2015,Scimiterna2017,Lobb2009}. In this article  we presented various equations defined on six and seven point stencils, which to the best of our knowledge are novel. The novelty can be inferred from the degrees on which the variables appear in the equations. Moreover these equations serve as non-potential versions of some known integrable equations defined on quads~\cite{ABS}. Note  that through our procedure we can also obtain multiquadratic, or even higher degree   6-point equations. We have omitted the presentation of these exotic systems here, however, they require further investigation. In addition, a challenging problem is to classify integrable six point equations and seven point equations.  

Finally, we have presented multiquadratic quad relations, as well as compatible equations of homogeneous degree defined on ten point stencils of the ${\mathbb Z^3}-$lattice, as well as various equations defined on ten and twelve point stencils. Equations defined on ten point stencils of the ${\mathbb Z^3}-$lattice have appeared rarely in the literature so far. The only example that we can point out is in \cite{Fu-Nijhoff-2017} where such an equation was introduced but without its compatible partner.


\end{document}